\documentclass[traditabstract]{aa} 
\usepackage{graphicx}
\usepackage{txfonts}
\usepackage{mathtools} 
\usepackage{numprint}
\usepackage{color}
\usepackage{placeins}
\npdecimalsign{.}
\nprounddigits{0}
\newcolumntype{M}[2]{>{\hfill}n{#1}{#2}<{\hfill}}
\newcolumntype{N}[2]{>{\hfill}n{#1}{#2}<{\hfill}}
\usepackage[colorlinks,linkcolor=blue,anchorcolor=blue,citecolor=blue]{hyperref}
\usepackage{newtxtext}
\usepackage[varvw]{newtxmath}
\usepackage{float}

\def\la{\lower.5ex\hbox{$\; \buildrel < \over \sim \;$}}
\def\ga{\lower.5ex\hbox{$\; \buildrel > \over \sim \;$}} 
\begin{document}
   \title{Predicting HCN, HCO$^{+}$, multi-transition CO, and dust emission of star-forming galaxies}

   \subtitle{The inner $30$~pc of the Circinus galaxy}

   \author{B.~Vollmer\inst{1}, D.~Iung\inst{1}, K.R.W.~Tristram\inst{2}, R.I.~Davies\inst{3}, W.M.~Goesaert\inst{4}}

   \institute{Universit\'e de Strasbourg, CNRS, Observatoire Astronomique de Strasbourg, UMR 7550, 67000 Strasbourg, France \and
     European Southern Observatory, Alonso de C\'{o}rdova 3107, Vitacura, Santiago, Chile \and
     Max-Planck-Institut für extraterrestrische Physik, Postfach 1312, 85741 Garching, Germany \and
     Leiden Observatory, University of Leiden, PO Box 9513, 2300 RA Leiden, The Netherlands
     }

   \date{Received ; accepted }

  \abstract
      {We present an analytical model of a turbulent clumpy star-forming disk that reproduces the spatially resolved observations for 6 molecular lines
        and 3 continuum bands at 250-700~GHz (0.4-1.2~mm) in the central 30 pc of the Circinus Galaxy.
        Using the properties of the stellar disk as input, and when matched to the line and continuum emission, the model yields the Toomre $Q$ parameter,
        mass accretion rate $\dot{M}$, metallicity $Z$, and cosmic ray ionization rate $\zeta_{CR}$.
        These in turn are used to infer the gas properties (surface density, velocity dispersion, and radial velocity) as well as the star formation rate.
        Our main findings from the model include a low CO-to-H$_2$ conversion factor that varies with radius in the range
        $\alpha_{\rm CO}=0.5$--1.5 $M_{\odot}$pc$^{-2}/({\rm K\,km\,s}^{-1})$, close to the canonical value for such environments.
        It shows that CO(3-2) is a poor tracer of the gas mass surface density $\Sigma_{{\rm H}_2}$ on these small scales, which is better represented by
        molecules such as HCO$^+$(3-2) with higher critical density.
        The steep relation between $\Sigma_{{\rm H}_2}$ and star formation rate surface density $\dot{\Sigma}_*$ is typical of compact starburst systems, and
        the star formation rate (10~$M_{\odot}$kpc$^{-2}$yr$^{-1}$ averaged over the disk) is high enough to drive a wind.
        Despite this, and in contrast to typical large scale galactic disks, the energy injection rate into the ISM (and hence the gas distribution
        within the disk) is dominated by gas inflow through the disk rather than the supernova rate.
        As such, the decreasing inflow rate at smaller radii could be caused by a wind, or indicate a recent increase of the mass inflow from larger
        ($>30$ pc) scales. We suggest that such variations in the inflow rate could be associated with the nested spiral structures which have been
        observed in Circinus.}

   \keywords{Galaxies: evolution; Galaxies: ISM; Galaxies: star formation; Radio lines: galaxies}

   \authorrunning{Vollmer et al.}

   \maketitle
   \nolinenumbers
%

\section{Introduction\label{sec:introduction}}

Supermassive Black Holes (SMBH) can grow through black hole merging or gas accretion. Theoretical studies and
observations suggest that major mergers are the dominant process leading to SMBH growth (Menci et al. 2014, Hopkins et al. 2014,
Gatti et al. 2015) for SMBH masses higher than $10^8$~M$_{\odot}$. SMBHs with masses lower than $10^7$~M$_{\odot}$ located
in disk galaxies mainly grow through secular processes (spirals and bars), which lead to gas inflows (Hopkins et al. 2014,
Davies et al. 2015, Rosario et al. 2018). For a review on the growth of SMBH, see Storchi-Bergmann \& Schnorr-M\"{u}ller (2019)
and Alexander et al. (2025).
With the advent of the Atacama Large Millimetric Array (ALMA), the mapping of gas flows on parsec scales is now possible \citep{alo18,alo19,imp19,tri22,goe25,oak25}.

At kpc-scales a galactic bar can efficiently transfer angular momentum within the disk, allowing gas to move inwards.
Since gas and stars are not aligned, stellar bars exert a torque on the gas, except at resonances,
where the gas configuration falls back into symmetry with the bar.
The infalling gas forms a ring at the Inner Lindblad Resonance (ILR; see e.g., Knapen et al. 2006) located at galactic radii
of a few hundred pcs to a kpc \citep{com10}. Between corotation and ILR, the gas spiral arms exert negative torques, and gas is driven inwards, increasing the depth of the potential well.  The vertical resonance with the bar develops a peanut-shape bulge,
weakening the bar. 
This can trigger the decoupling of a secondary bar, an embedded nuclear bar, rotating faster than the primary bar (e.g. Buta and Combes, 1996).
However, high resolution hydrodynamical simulations of gas flows in barred galaxies have shown that a secondary bar can prevent rather than enhance further inflow because it ends well inside its corotation radius \citep{mac00,mac02}.
The large scale bar may instead drive a circumnuclear spiral which can drive gas to smaller scales \citep{mac04a,mac04b}.
\citet{mar03a} showed that circumnuclear dust structures span a range of morphologies - spiral, chaotic, or absent - in both barred and unbarred galaxies, but found no cases with secondary bars.
Notably, grand design circumnuclear spirals were found almost exclusively in barred galaxies (\citealt{mar03b}; see also \citealt{dav14}), supporting the theoretical understanding that these can bring the gas down to 10--100~pc scales.
Multiscale simulations \citep{hop10,hop24} show that one can expect such structures to be repeated on smaller scales in a nested hierarchy, with the details depending either on the initial cosmological environment, or on a variety of galaxy properties such as prominence of the bulge, mass of the black hole, and gas fraction.
In this context, it is no surprise that disks and their associated structures are observed on a variety of increasingly smaller spatial scales.

High-resolution ALMA observations of the central $\sim 100$~pc regions of disk galaxies reveal distinct circumnuclear structures that regulate
nuclear fueling. For example, Combes et al. (2019) identified resonant gas rings on a $100$~pc scale within seven active spiral galaxies, which serve as
reservoirs feeding the nucleus, occasionally displaying inner trailing spirals and decoupled circumnuclear disks (see also Audibert et al. 2021)
with radii of $6$–$30$~pc and masses of $\sim 10^7$~M$_{\odot}$. Similarly, Garcia-Burillo et al. (2021) resolved dusty molecular disks around the AGN
of ten nearby Seyfert galaxies via $870~\mu$m continuum and CO line emissions. These disks are characterized by a median diameter of $\sim 40$~pc,
a median molecular gas mass of $\sim 6 \times 10^5$~M$_{\odot}$, and an orientation perpendicular to the AGN wind axes. The gas masses are uncertain because
of the unknown CO-to-H$_2$ conversion factor, especially for higher-transition CO lines.

The Circinus galaxy hosts the nearest type~2 Seyfert active galactic nucleus (AGN) at a distance of $4.2$~Mpc ($0.1''=2$~pc; Tully et al. 2009).
Because it is so near, spiral structures have been observed to exist on several distinct scales in the central regions.
They extend across radii from several kpc down to $\sim2$~pc over three or four hierarchical levels, and are all oriented in the same direction
with trailing arms.

Recent APEX and ALMA multi-scale molecular line observations of the Circinus galaxy revealed a complex, hierarchical structure.
At the kiloparsec scale, Lasrado et al. (2025) identified a molecular gas bar characterized by an X-shaped position-velocity diagram indicative
of a weak bar potential, while Zschaechner et al. (2016) mapped a symmetric main disk with extended spiral arms alongside an inner bar structure
($\sim 200$~pc) and a significant molecular outflow ($0.35$–$12.3$~M$_{\odot}$yr$^{-1}$) originating from the central gas disk ($R \le 80$~pc).
Probing closer to the nucleus, Izumi et al. (2018) resolved this nuclear environment into gas arms and a circumnuclear disk (CND; $74$~pc $\times$ $34$~pc)
with a molecular mass of $M_{\rm H_2} \sim 3 \times 10^6$~M$_{\odot}$, which Tristram et al. (2022) further characterized as a geometrically thin,
rotation-dominated structure harboring a compact, unresolved torus ($\la 4$~pc) traced by high-density molecular emission. At the highest angular
resolution ($0.25$~pc), Goesaert et al. (2025) demonstrated that the CND hosts molecular and dusty spiral arms that channel gas inward toward the
torus at a feeding rate of $0.3$--$7.5$~M$_{\odot}$yr$^{-1}$ via a non-axisymmetric potential; because this inflow drastically exceeds the supermassive
black hole accretion rate of $\la 0.01$~M$_{\odot}$yr$^{-1}$, the majority of this transported mass is ultimately expelled in a multi-phase outflow
before reaching the accretion disk.

The emerging picture from these molecular line observations is thus the following: at kpc-scales, gas is driven inward by a
weak large-scale bar. The inflowing gas forms spiral arms that feed a relatively symmetric CND in the inner $R \sim 30$~pc of
the Circinus galaxy.

In this work, we study the properties of the CND and its ability to form stars.
To do so, we applied the analytical model for the molecular line and continuum emission of galactic disks presented in Vollmer et al.
(2017, 2022, and 2025) to the observations of Tristram et al. (2022) and Goesaert et al. (2025). The comparison between the model
molecular line and continuum observations yields
the H$_2$ gas mass, gas velocity dispersion, Toomre $Q$ parameter, and star formation rate of the gas disk. The Circinus CND ($R \sim 30$~pc)
can then be compared to galactic other disks at kpc scales.

This article is structured in the following way: the model is briefly described in Sect.~\ref{sec:model}, the data in Sect.~\ref{sec:data}.
The model results for the radial profiles and images are presented in Sect.~\ref{sec:modelresults}, followed by the derived physical parameters
(Sect.~\ref{sec:parameters}). 
These results are set into context and discussed in Sect.~\ref{sec:discussion}. Finally, we give our conclusions in
Sect.~\ref{sec:conclusions}.

\section{Model \label{sec:model}}

We use the one-dimensional analytical model described in Vollmer et al. (2017, 2025) where galactic gas disks are modeled as turbulent
clumpy star-forming accretion disks.
This model was also used in Vollmer et al. (2022, 2025) to calculate the integrated radio continuum emission of the galaxies.
The model considers the warm, cool neutral, and molecular phases of the interstellar medium as a single, turbulent gas
in vertical hydrostatic equilibrium. The details of the large-scale gas disk model are described in Appendix~\ref{sec:dgasdisk}.
The model inputs are the observed rotation curve and stellar surface density. The stellar velocity dispersion has to be assumed.
These parameters define the gravitational potential of the disk. Our model approach does not account for strong non-circular motions.
Therefore, we apply our model only to the Circinus CND, since it appears as the inner extension of larger-scale spiral arms, where the gas motions circularize.

The gas is assumed to be clumpy, so that the local density is enhanced relative to the average density of the disk. 
Using the local density, the free fall time $t_{\rm l,ff}$ of an individual cloud together with the volume filling factor of
self-gravitating clouds control the timescale for star formation on scales $\ga 100$~pc (Eq~\ref{eq:star-form}).
This prescription leads to a star formation surface density that is proportional to the turbulent gas pressure (see also Ostriker \& Kim 2022).

The turbulence of the interstellar medium (ISM) is taken into account by
applying scaling relations for the gas density and velocity dispersion to gas clouds of given sizes. The scaling relations change when the clouds become self-gravitating: from those found by Quiroga et al. (1983) for galactic H{\sc i} to that of Larson et al. (1981) and Lombardi et al. (2010)
for selfgravitating clouds.
The cloud temperature was calculated via the balance of heating and cooling. In this way, the gas density, temperature, and
velocity dispersion were determined for each cloud. The molecular abundances depend on the density, temperature,
and cosmic-ray (CR) ionization rate, which was calculated by using a steady-state model calibrated by the CR ionization in the
solar neighborhood. The molecular abundances, gas density, velocity dispersion, and temperature served as input for the calculation of the molecular line emission.
The radio continuum emission was calculated via a steady-state model calibrated by the observed IR-radio correlations (Vollmer et al. 2025).

The model simultaneously calculates radial profiles of the gas surface density $\Sigma$, 
the SF surface density $\dot{\Sigma}_*$, the gas velocity dispersion $v_{\rm turb}$, and the volume-filling factor $\Phi_{\rm v}$, all of which are large-scale properties,
and at small scales $f_{\rm mol}$ and the IR SED and molecular line emission.  
Each ring at a given radius is handled independently given its rotation velocity, epicyclic frequency, stellar mass surface density and velocity dispersion.
The molecular line emission was calculated via RADEX (van der Tak 2007).
For the details of our analytical model, we refer to Appendix~A of Liz\'ee et al. (2022).
The model has been successfully applied to reproduce the IR spectra energy distributions, CO, HCN, and HCO$^+$ line luminosities,
and the CO spectral line energy distributions of main-sequence and starburst galaxies across redshifts $z=0$ to $z \sim 1$–$2$
(Vollmer et al. 2017, 2025). CO(1-0) and HCN(1-0) conversion factors were determined for all galaxy samples.

Our model can also be applied to two-dimensional maps as long as the velocity field, stellar mass distribution and velocity dispersion are approximately the
same in a ring. 
Since the Circinus CND appears as the inner extension of larger-scale spiral arms, where the gas motions circularize (Fig.~14 of Izumi et al. 2018;
Fig.~9 of Tristram et al. 2022), we apply our model to the radial profiles and maps only within the CND at $R < 30$~pc.

The rotation curve $v_{\rm rot}$, the stellar surface density $\Sigma_*$, and stellar velocity dispersion $v^*_{\rm disp}$ profiles served as model
  inputs (Table~\ref{tab:mparams}). We adopted the rotation curve found by Tristram et al. (2017), together with a black hole mass of
  $M_{\rm BH}=2 \times 10^6$~M$_{\odot}$, which is consistent with the mass estimate of Greenhill et al. (2003).
  For the stellar mass distribution within a radius of $30$~pc we assume an exponential nuclear disk with a central surface density of
  $\Sigma_*=1.6 \times 10^4$~M$_{\odot}$pc$^{-2}$ and a scale length of $R_0=40$~pc. This mass distribution is consistent with the outer rotation curve
  (Fig.~\ref{fig:rotcurve}).
  In our fiducial model we added a second exponential stellar disk with $\Sigma_*=10^5$~M$_{\odot}$pc$^{-2}$ and $R_0=4$~pc. The scale length is consistent with the
  effective radius of the nuclear stellar cluster or disk observed by Pechetti et al. (2020). The resulting mass distribution is consistent with
  the rotation curve. Since the gas to stellar mass ratio is very small (Sect.~\ref{sec:ISM}), the gravitational potential is dominated by the stars.
  For simplicity, we kept the stellar velocity dispersion constant $v_{\rm disp,*}=40$~km\,s$^{-1}$.

  In a second model approach we replaced the latter stellar disk by a nuclear star cluster (see Sect.~\ref{sec:NSC}).
  In the following, we only show the fiducial model because it lead to significantly lower $\chi^2$ values at radii smaller than $10$~pc than those of the
  model including a nuclear star cluster.

The model contains five free parameters (Table~\ref{tab:mparams}):
(i) the Toomre parameter $Q=\frac{c_s \kappa}{\pi G \Sigma}$ of the gas, where a $Q<1$ disk is gravitationally unstable
(ii) the mass accretion rate $\dot{M}$, where an increasing $\dot{M}$ at constant $Q$ increases $v_{\rm turb}$, (iii) $\delta = l_{\rm driv}/l_{\rm cl}$,
where $l_{\rm cl}$ is the cloud size scale and $l_{\rm driv}$ is the driving length of the turbulence, (iv) the normalization of the gas metallicity $k_{\rm Z}$,
(Eq.~\ref{eq:metulirgs}) and (v) the normalization of the CR injection rate with respect to the SF volume density $k_{\rm CR}$ (see Sect.~2 of Vollmer et al. 2025),
which influences the gas chemistry and thus the molecular abundances.
The normalization $k_{\rm CR}=1$ is that found for main sequence and starburst galaxies at various redshifts by Vollmer et al. (2026).
  It implies that the CR ionization rate is higher by a factor of three than the rate determined for the solar neighborhood.
  The normalization $k_{\rm C}=1$ (see Eq.~\ref{eq:zzodot}) corresponds to the normalization of the metallicity with respect to the gas fraction of
  Vollmer et al. (2026).
  In agreement with Vollmer et al. (2017, 2022, 2025, 2026) and Liz\'ee et al. (2022), we set $\delta=5$.
  This leaves four open parameters for our model calculations: $Q$ and $\dot{M}$ set the properties of the gas disk (see Sect.~A.2 of Vollmer \& Leroy 2011),
  $k_{\rm Z}$ and $k_{\rm CR}$ set the molecular abundances and thus influence the IR and molecular line emission (see Vollmer et al. 2017 and 2025).
For each radius $R$ with a given rotation velocity, stellar mass surface density, and stellar velocity dispersion the
model yields the SF surface density $\dot{\Sigma}_{*}$, the H{\sc i} emission, and the IR continuum and molecular line emission for a given set of
$Q$, $\dot{M}$, $k_{\rm Z}$, and $k_{\rm CR}$.  
The constant that links the rate of energy injection by supernovae to the SFR was set at $\xi = 9.2 \times 10^{-8}$~(pc\,yr$^{-1}$)$^2$ (Vollmer et al. 2025, 2026).
Moreover, the dust had an assumed constant mass fraction (1\,\% of the hydrogen mass; see, e.g., Draine et al. 2007).

\begin{table}
      \caption{Model inputs, outputs, and free parameters}
         \label{tab:mparams}
      \[
       \begin{tabular}{ll}
         \hline
         input parameters \\
         \hline
         $v_{\rm rot}$ & rotation curve$^{\rm a}$ \\
         $\Sigma_*$ & stellar surface density$^{\rm a}$ \\
         $v_{\rm disp}^*$ & stellar velocity dispersion$^{\rm b}$ \\
         $\delta=l_{\rm driv}/l_{\rm cl}=5$ & size of selfgravitating clouds$^{\rm b}$\\
          \hline
          outputs$^{\rm c}$ calculated by& the model \\
           \hline
         $\rho$ & gas density\\
         $\Sigma$ & gas surface density\\
         $H$ & disk height\\
         $v_{\rm turb}$ & turbulent velocity dispersion\\
         $l_{\rm driv}$ & turbulent driving length scale\\
         $\dot{\rho}_*$ & SF volume density\\
           $\dot{\Sigma}_*=\dot{\rho}_*\,l_{\rm driv}$ & SF surface density\\
           $I_{\lambda}$ & IR spectral energy distribution\\
         $I_{\rm mol}$ & spectral line energy distribution\\
         \hline
         free parameters & (model inputs)$^{\rm d}$  \\
         \hline
         $Q$ & Toomre parameter\\
         $\dot{M}$ & mass accretion rate\\
         $k_{\rm Z}$ & metallicity normalization\\
         $k_{\rm CR}$ & CR injection rate normalization\\
          \hline
        \end{tabular}
       \]
       \begin{list}{}{}
       \item
         $^{\rm a}$ derived from observations; $^{\rm b}$ assumed;
         $^{\rm c}$ see Appendix~\ref{sec:dgasdisk} and Sect.~A.2 of Vollmer \& Leroy (2011) for the dependencies
         of the output parameters on the input parameters;
         $^{\rm d}$ fitted for by matching the observed IR SED and line emission.
       \end{list}
\end{table}

We calculated model grids of the four open parameters and searched for the models that best matched the observed continuum and molecular line measurements.
To do so we calculated the $\chi^2$ between the model and observed continuum and line measurements where we used the scatter in an annulus of radius
$R$ (following Leroy et al. 2008) as the uncertainty of the measurement for the $\chi^2$.
The total $\chi^2$ for a given radius is the sum of the individual $\chi^2$.
When applying the model to the 2D data, the comparison to the model grid is carried out using the molecular line maps together with the band~9
continuum map only, since the latter is least affected by free–free emission. For this comparison, the rms of the line-free channels in each voxel
(for the lines) and the rms of the band~9 map are adopted as uncertainties in the $\chi^2$ calculation.
In total, $2640$ models with different values of $Q$, $\dot{M}$, $k_{\rm Z}$, and $k_{\rm CR}$ were calculated per galactic radius. 
To obtain the mean value and uncertainties of the derived parameters, we took the
$10$ models with the lowest $\chi^2$ into account, which corresponds to an upper limit of $\chi_{\rm lim}^2 \sim 2$.
We call these models the best-fit models.
In this way the mean physical parameters of the gas disk (mass surface density, volume density, velocity dispersion, turbulent driving scale length,
disk height) and their uncertainties could be recovered.

\subsection{Cosmic-ray ionization rate \label{sec:crion}}

Cosmic rays regulate the chemistry of the ISM through ionization, affecting molecular abundances and thus the predicted line emission
(e.g. Dalgarno 2006, Tielens 2013).
As in Vollmer et al. (2025) we calculated the CR ionization rate, which was then injected into 
the time-dependent gas-grain code Nautilus, which uses the total H$_2$ CR ionization rate $\zeta_{{\rm H}_2}$ as input (Wakelam et al. 2012).
The CR nucleons (mainly protons) lose energy by the combined effects of ionization and Coulomb losses, adiabatic
deceleration, and pion production. To calculate the local (solar neighborhood) CR density spectra of electrons and protons,
we followed Pohl (1993). The normalization factor of the proton spectrum $k_{\rm CR}$ was chosen such that the observed local CR intensity,
averaged over a solid angle, $J(E)$, was recovered (Fig.~3 of Vollmer et al. 2025).
To calculate the CR ionization rate at a given distance $R$ from the center of a galaxy, we set $n_{\rm H}$ to the
gas density of the analytical galaxy model (Appendix~\ref{sec:dgasdisk}), use the interstellar radiation field $U$ and the magnetic field strength
of the galaxy model at $R$, and multiply the proton and electron spectra by $\dot{\rho}_*$ of the galaxy model.
In the Vollmer et al. (2025) the normalization was done with the SF surface density $\dot{\Sigma}_*$.
The new normalization $k_{\rm CR}$ leads to significantly higher CR ionization rates in starburst galaxies. However, this only leads to
a decrease of the molecular line emission of these galaxies by $\la 50$\,\%.

\subsection{Model grids}

For the direct comparison with the available continuum and line observations, we calculated a grid of models by varying
the free parameters $Q$, $\dot{M}$, the normalization of the CR injection rate $k_{\rm CR}$ with respect to the local value, the normalization of the gas
metallicity $k_{\rm Z}$. The ranges of these four parameters are: $1 \le Q \le 10$, $10^{-2} \le \dot{M} \le 10$~M$_{\odot}$yr$^{-1}$,
$k_{\rm CR}=0.1,0.3,0.5,1,2,3$, and the normalization of the gas metallicity $k_{\rm Z}=0.3,0.5,1,2,3$.

\section{Data \label{sec:data}}

A list of the available data is shown in Table~\ref{tab:data}.
\begin{table}
      \caption{Line and continuum data}
         \label{tab:data}
      \[
       \begin{tabular}{lcccc}
         \hline
         Line or & frequency & beam & PA$_{\rm beam}$ & ref. \\
         continuum & (GHz) & (mas) & (degrees) & \\
         \hline
         Low resolution & & & & \\
         \hline
         B6LR cont. & 258.8 & 127 $\times$ 102 & -31 & (2) \\
         B7 cont. & 345.4 & 190 $\times$ 120 & 27 & (1) \\
         HCO$^+$(3--2)LR & 267.6 & 127 $\times$ 102 & -31 & (2) \\
         HCN(3--2)LR & 272.0 & 127 $\times$ 102 & -31 & (2) \\
         CO(3--2) & 345.8 & 190 $\times$ 130 & 26 & (1) \\
         HCO$^+$(4--3) & 356.7 & 180 $\times$ 120 & 27 & (1) \\
         HCN(4--3)$^{\rm (a)}$ & 354.5 & 190 $\times$ 120 & 31 & (1) \\
         \hline
         High resolution & & & & \\
         \hline
         B6HR cont. & 258.8 & 22 $\times$ 19 & 27 & (2) \\
         B7 cont. & 349.6 & 49 $\times$ 29 & 18 & (2) \\
         B9 cont. & 691.5 & 110 $\times$ 70 & 42 & (1) \\
         HCO$^+$(3--2)HR & 267.6 & 22 $\times$ 19 & 27 & (2) \\
         HCN(3--2)HR & 272.0 &22 $\times$ 19 & 27 & (2) \\
         CO(3--2) & 345.8 & 49 $\times$ 29 & 18 & (2) \\
         CO(6--5) & 691.5 & 100 $\times$ 70 & 34 & (1) \\
         HCO$^+$(4--3) & 356.7 & 49 $\times$ 29 & 18 & (2) \\
         HCN(4--3) & 354.5 & 49 $\times$ 29 & 18 & (2) \\
         \hline
       \end{tabular}
        \]
       \begin{list}{}{}
       \item (1) Tristram et al. (2022); (2) Goesaert et al. (2025). \\
         $^{\rm (a)}$ The HCN(4--3) line was positioned at the edge of
         its spectral window and only the blue-shifted part of the line up
         to the systemic velocity could be covered.
       \end{list}
\end{table}
We used the ALMA band~7 and 9 observations from Tristram et al. (2022), which include CO(3–2), CO(6–5), HCN(4–3), and HCO$^+$(4–3) at $\sim 100$~mas
resolution, as well as the band~6 and 7 data at $20$–$50$~mas resolution from Goesaert et al. (2025), additionally including HCN(3–2) and HCO$^+$(3–2).
Additional low-resolution band~7 CO(3–2) data from Goesaert et al. (2025) are not used, as the line lies at the edge of the band and is affected by artifacts.

We realized that the Tristram et al. (2022) CO(3-2) moment~0 map has significantly lower fluxes in the inner $5$~pc compared to the
low-resolution CO(3-2) moment~0 map of Goesaert et al. (2025). We can only speculate that Tristram et al. (2022) oversubtracted the
continuum emission in this region. After some testing, we opted to convolve and regrid the high-resolution cube to match the low-resolution
cube. Furthermore, a binary mask based on a $3\sigma$ clipping of the low-resolution cube was applied to the high-resolution cube and
moment~0 maps were calculated from both data cubes.
We then took the higher flux from the Tristram et al. (2022) and Goesaert et al. (2025) maps for each pixel of the final moment~0 map.
This procedure increased the CO(3-2) radial profile for $R \le 5$~pc where there is little resolved-out flux in the high-resolution data
because its maximum recoverable scale is $0.6''$ or $\sim 12$~pc (Goesaert et al. 2025).
The CO(3-2) profile for $R > 5$~pc is essentially that of the low-resolution data.
A comparison to the $290 \times 240$~mas CO(3-2) map of Izumi et al. (2018) showed that the flux within the inner $5$~pc recovered by our
final map is still somewhat lower but comparable that of Izumi et al. (2018). It turned out that model results for $R \le 5$~pc
do not change if the CO(3-2) line emission is dismissed from the modeling. However, the CO(3-2) line emission is important for the
model results at larger radii.

To compare the data to the model we produced two datasets: (i) the low-resolution (LR) dataset with a common resolution of $200$~mas given by the
Tristram et al. (2022) band~7 observations; the resolutions of the band~9 and CO(6-5) maps were degraded to match those of the band~7 image and
(ii) a high-resolution (HR) dataset with a common resolution of $100$~mas given by the
Tristram et al. (2022) band~9 observations. Both datasets contain the band~6, 7, 9 continuum images and the CO(3-2), CO(6-5), HCN(3-2), HCO$^+$(3-2),
HCN(4-3), and HCO$^+$(4-3) moment~0 maps. From these images we extracted radial profiles assuming an inclination of $i=70^{\circ}$ and a position
angle of PA$=220^{\circ}$. 
Since the HCN(4-3) line was positioned at the edge of its spectral window in the LR dataset and only the blue-shifted part of the line up to the systemic
velocity could be covered (Tristram et al. 2022), the radial profiles of the LR data were extracted from the eastern half of the moment~0 maps.

\section{Model results \label{sec:modelresults}}

In this Section we first assess if the model is a good fit to the data. We then have a look in more detail at the comparison between the model and observed
radial profiles and maps. The LR and HR models were calculated independently.

\subsection{Radial profiles}

\begin{figure}
  \centering 
  \resizebox{\hsize}{!}{\includegraphics{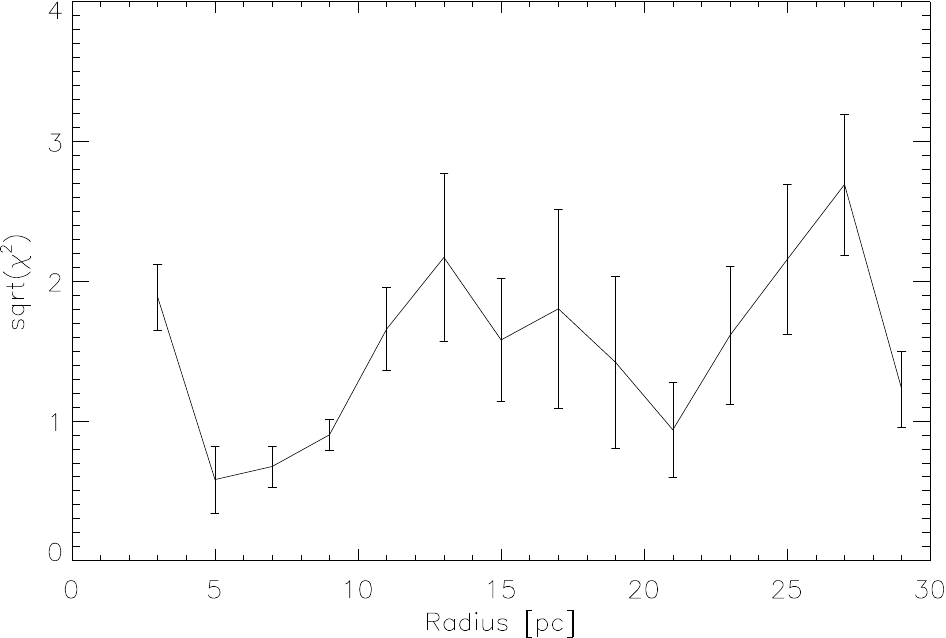}}
  \resizebox{\hsize}{!}{\includegraphics{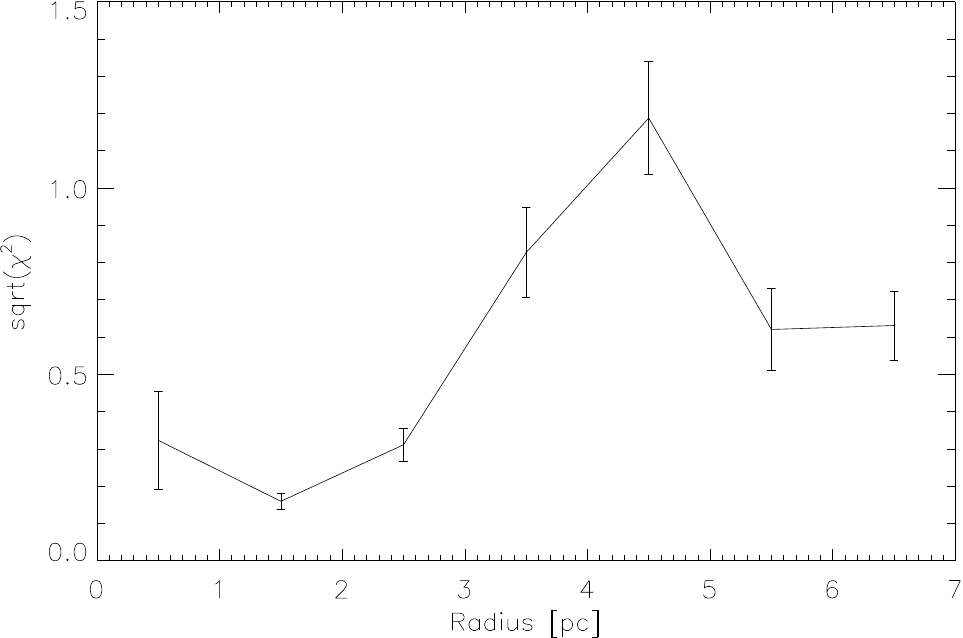}}
  \caption{Mean $\chi^2$ for the $10$ best models and its scatter. Upper panel: LR model, lower panel: HR model.
  \label{fig:khideux_log_nd_pb6_pco32HR_konrad_nooutflow_2disks_18}}
\end{figure}

The mean $\chi^2$ for the $10$ best models and its scatter are presented in
Fig.~\ref{fig:khideux_log_nd_pb6_pco32HR_konrad_nooutflow_2disks_18} for the LR and HR datasets, respectively.
Since large-scale structures are not well recovered by the HR datasets, we restrict the radial range to $6.5$~pc for the HR datasets.
The $\sqrt{\chi^2}$ values of the LR models are smaller than $2.5$, whereas those of the HR model are smaller than $1.2$.
The $\chi^2$ values of the HR model smaller than unity are due to a slight overestimation of the uncertainties of the observed radial profiles.
Whereas the $\sqrt{\chi^2}$ values of the LR and HR models at $5$ and $7$~pc are the same, those at $R=3$~pc differ significantly.
The $\sqrt{\chi^2}$ value of the HR model is lower than that of the LR model. We conclude that the LR and HR models
reproduce the available observations in a satisfactory way with $\sqrt{\chi^2} \la 2$.

\begin{figure}
  \centering
  \resizebox{\hsize}{!}{\includegraphics{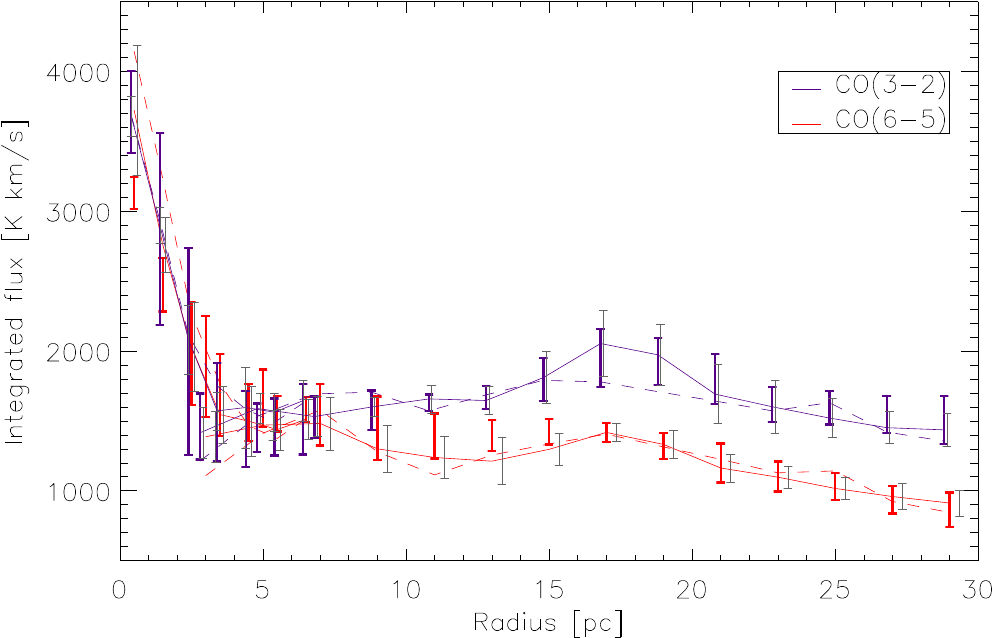}}
      \resizebox{\hsize}{!}{\includegraphics{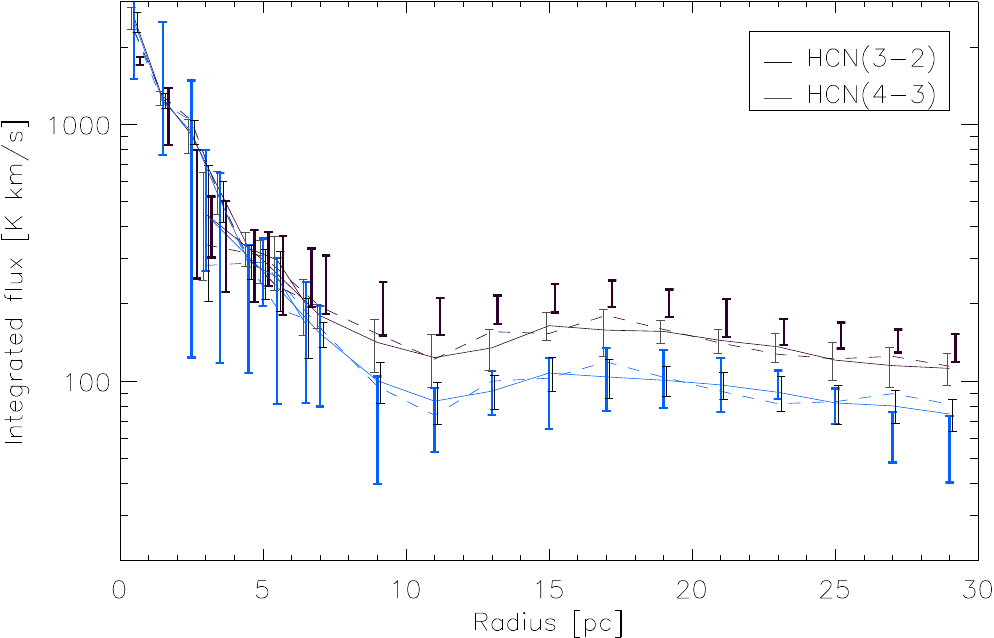}}
  \resizebox{\hsize}{!}{\includegraphics{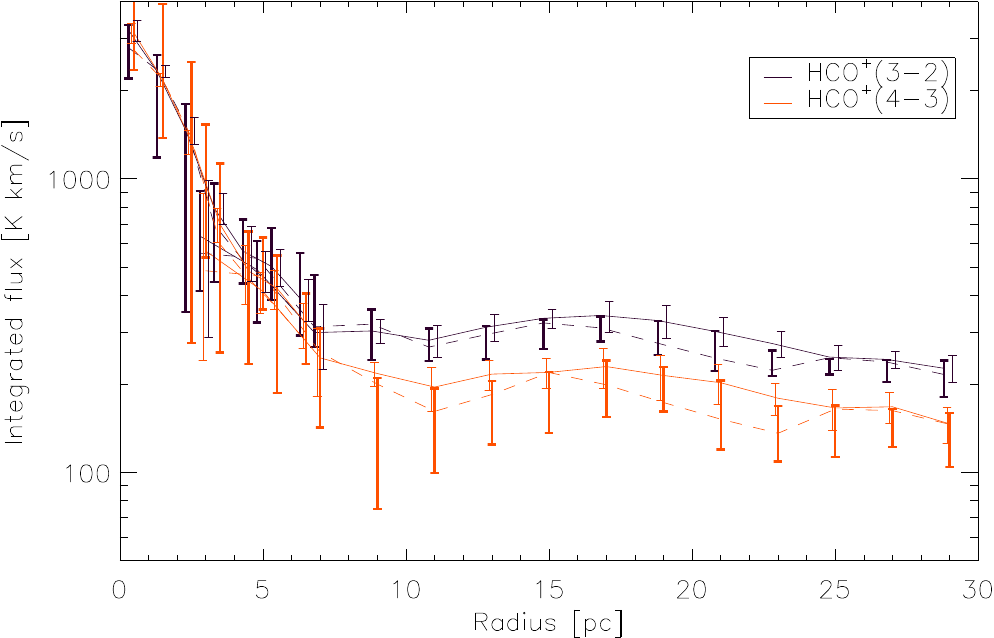}}
  \caption{Comparison between the observed and model radial profiles of the molecular line emission.
    The HR ($1{\ \rm pc} \le R \le 7{\ \rm pc}$ and LR ($R \ge 3$~pc) models are shown together.
    Solid line: mean of the $10$ models with the lowest $\chi^2$.
    Thin error bars: mean of the $10$ best-fit models. Dashed line: best-fit model; thick error bars: observations. 
  \label{fig:fig2}}
\end{figure}

The comparison between the observed and model radial profiles of the molecular line emission is shown in Fig.~\ref{fig:fig2}.
The HR data of all molecular lines are reproduced by the model within the error bars at all radii.
The CO(3-2), CO(6-5), and HCN(4-3) LR data are also reproduced by the model within the error bars at all radii.
However, the model somewhat underestimated the CO(6-5) LR flux between $7$ and $15$~pc.
The HCN(3-2) LR data are underestimated by the model by less than a factor of two.
The HCO$^{+}$(3-2) and  HCO$^{+}$(4-3) LR models are located at the upper limit of the associated data.
We conclude that the LR and HR model radial profiles of the molecular line emission reproduce the available observations in a satisfactory way.

\begin{figure}
  \centering
  \resizebox{\hsize}{!}{\includegraphics{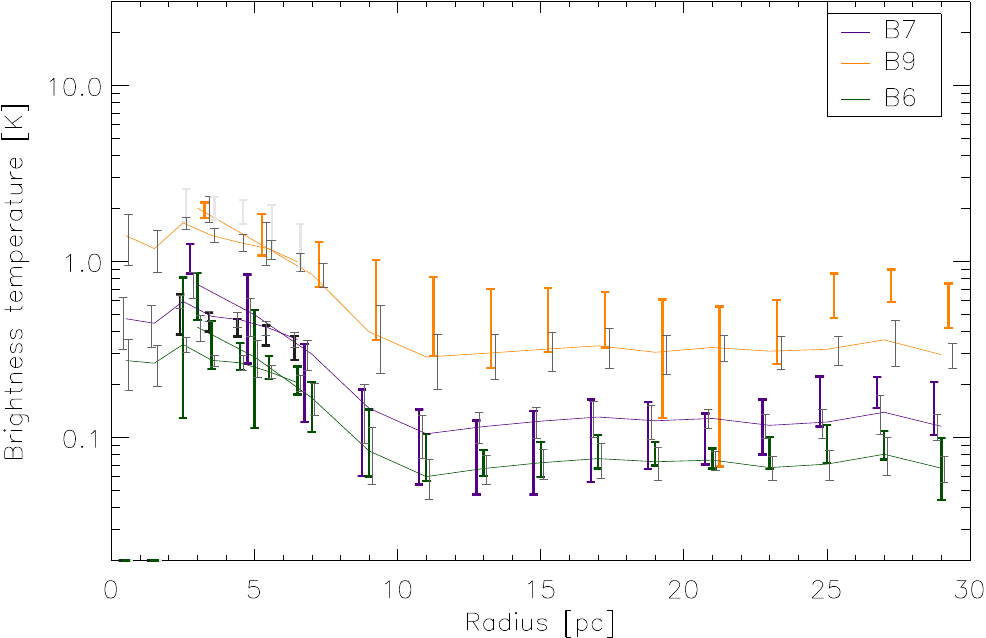}}
  \caption{Comparison between the observed and model radial profiles of the continuum emission.
    The HR ($1{\ \rm pc} \le R \le 7{\ \rm pc}$ and LR ($R \ge 3$~pc) models are shown together.
    Thin error bars: mean of the $10$ best-fit models. Thick error bars: observations. Upper panel: LR model; lower panel: HR model.
  \label{fig:fig5}}
\end{figure}

The comparison between the observed and model radial profiles of the continuum emission is shown in Fig.~\ref{fig:fig5}.
As shown by Goesaert et al. (2025), the continuum emission in the inner $3$~pc is dominated by free-free emission from gas, which is
directly illuminated by the UV emission of the AGN. We therefore decided to mask this region when the radial profiles were extracted.
We tried a masking, where the continuum emission was set to zero within a given radius.
However, the free-free emission could not be removed in a satisfactory way for any radius.
The best result was obtained by a vertical rectangular masks with a extension in RA of $4$~pc and $3$~pc for the LR and HR data, respectively.
We note that there are no observed HR continuum fluxes at radii $R < 2$~pc. For these radii the $\chi^2$ is only calculated from the
molecular lines. Hence, the continuum emission was predicted for these radii.

The band~6 and 7 LR and HR data are reproduced by the model within the error bars at all radii.
The band~9 LR and HR data are underestimated by the model by less than a factor of two.
We conclude that the LR and HR model radial profiles of the continuum emission reproduce the available observations in a satisfactory way.

\subsection{Moment~0 maps}

\begin{figure*}
  \centering
  \resizebox{\hsize}{!}{\includegraphics{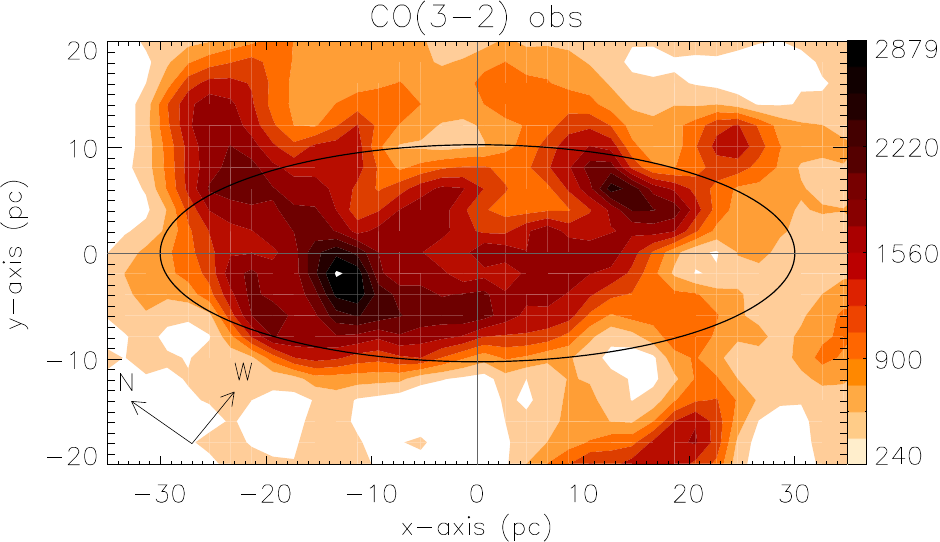}\includegraphics{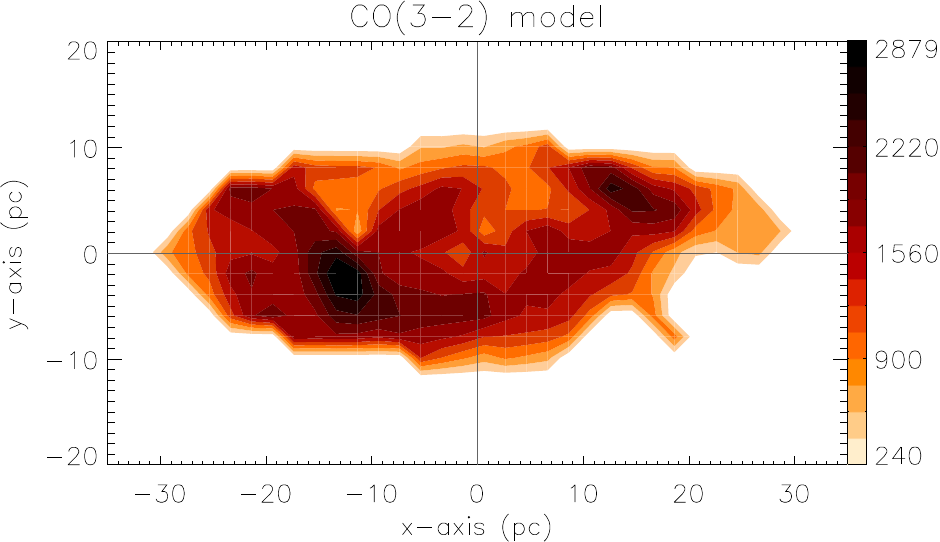}\includegraphics{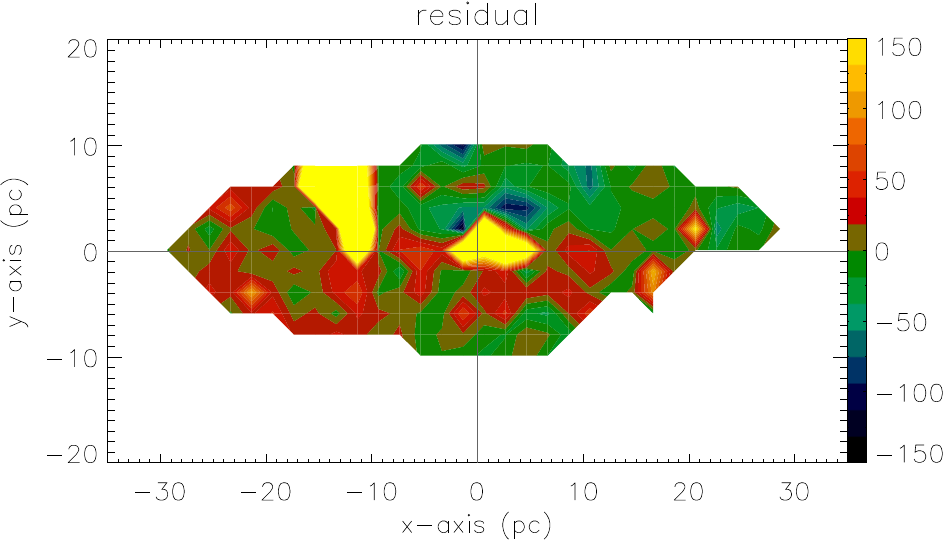}}
  \resizebox{\hsize}{!}{\includegraphics{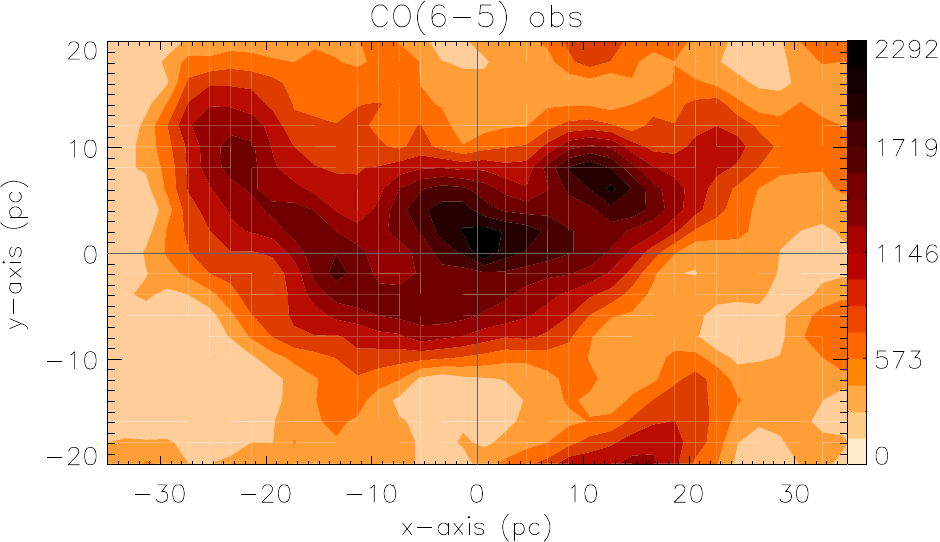}\includegraphics{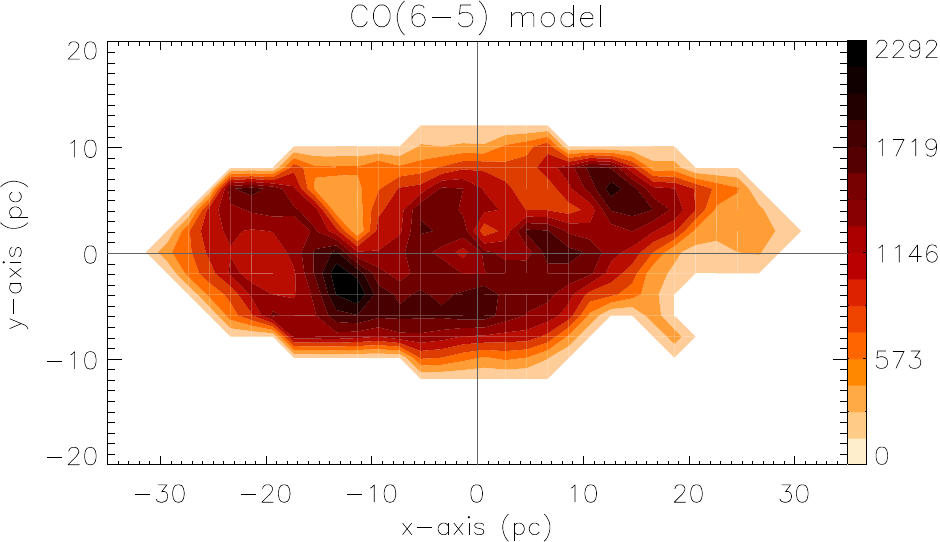}\includegraphics{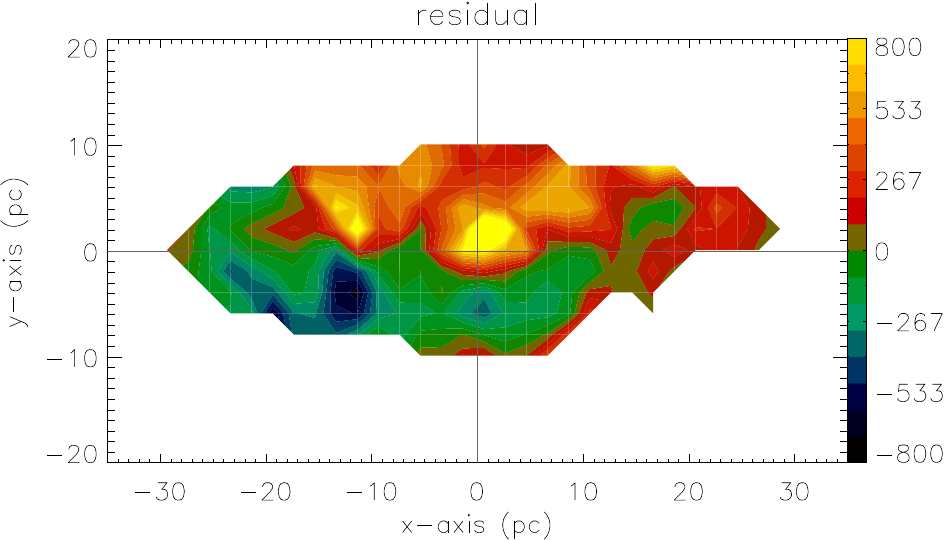}}
  \resizebox{\hsize}{!}{\includegraphics{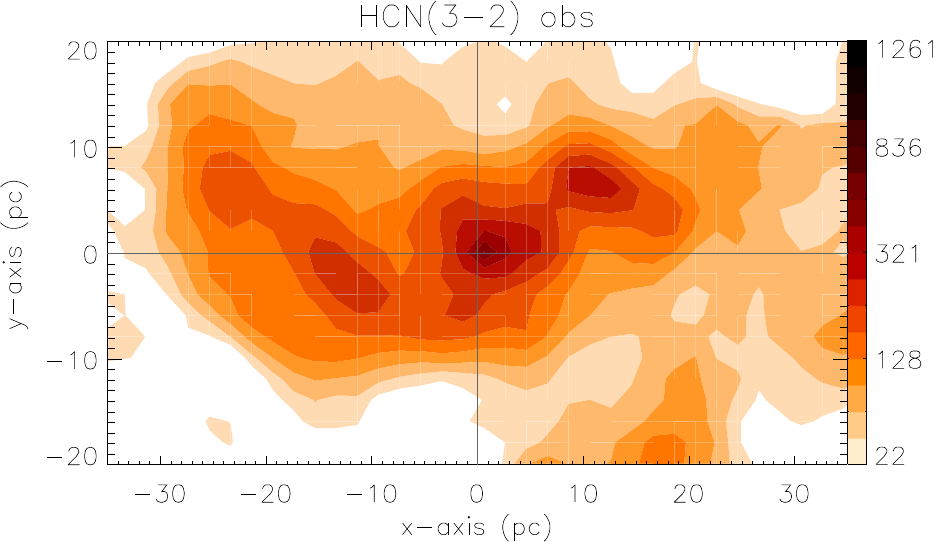}\includegraphics{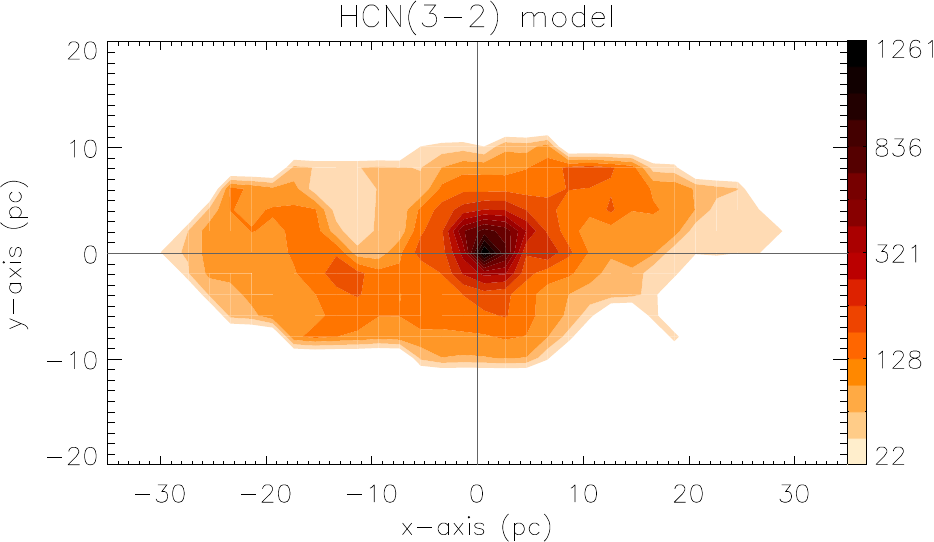}\includegraphics{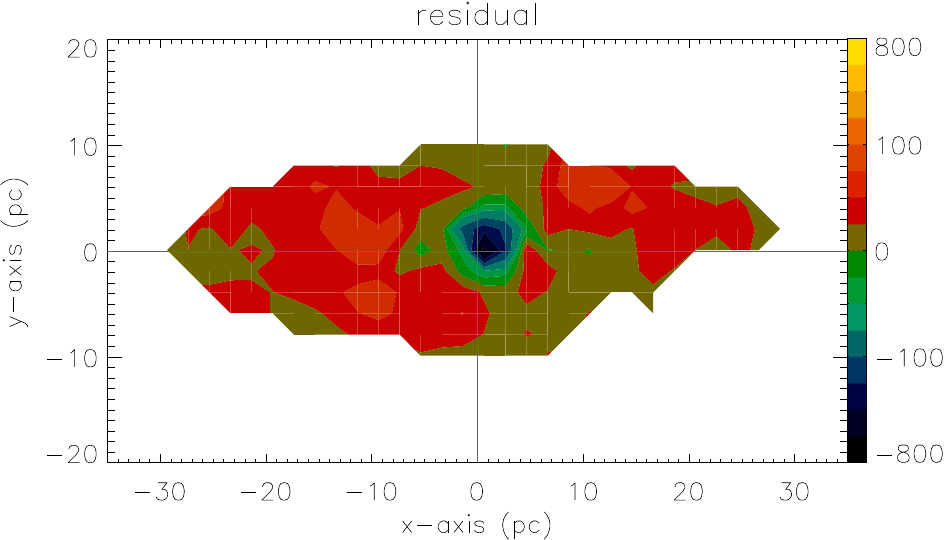}}
  \resizebox{\hsize}{!}{\includegraphics{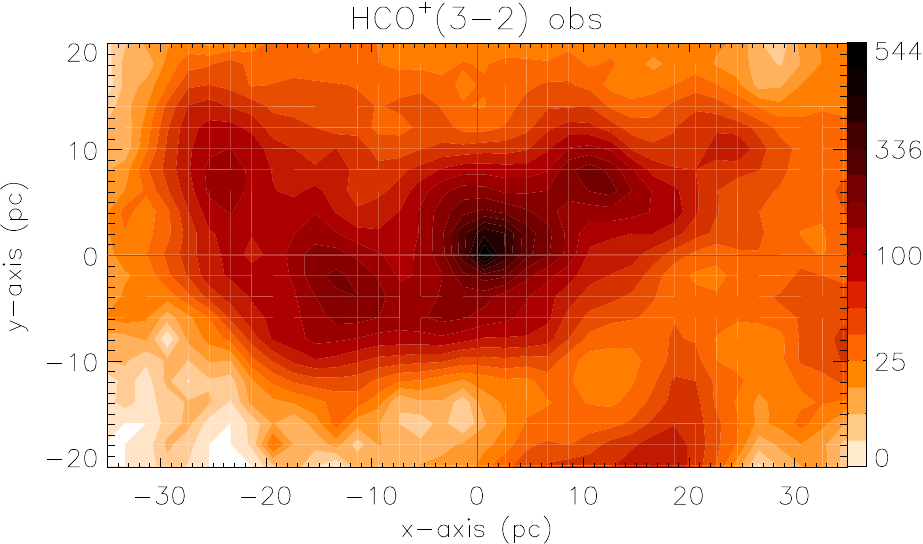}\includegraphics{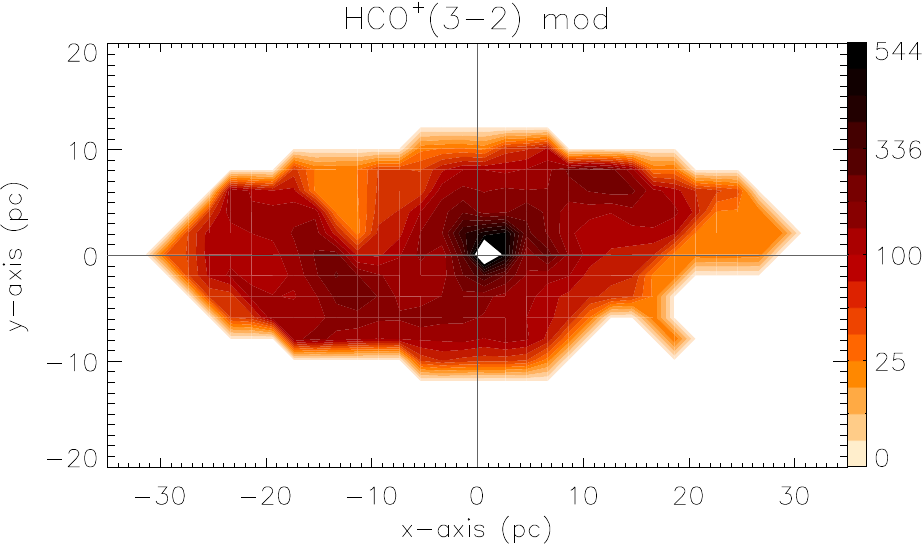}\includegraphics{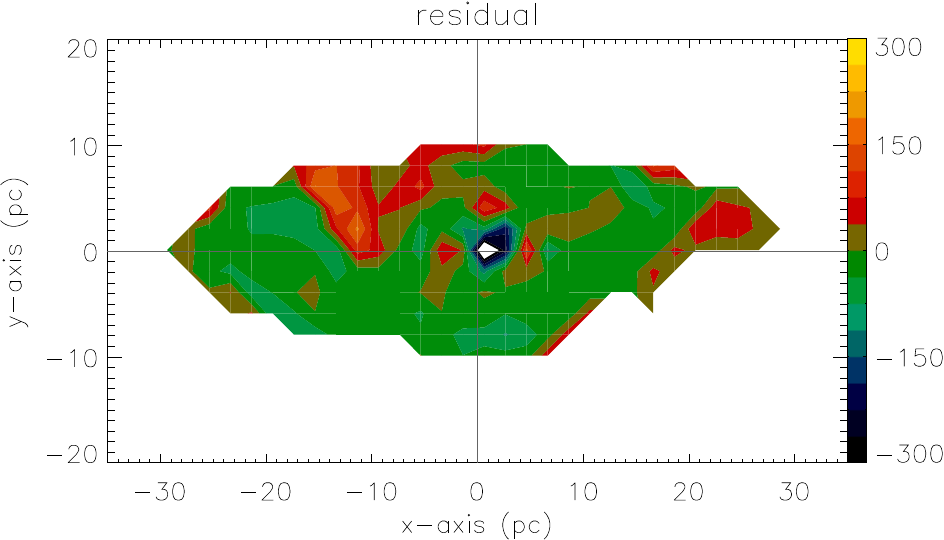}}
  \resizebox{\hsize}{!}{\includegraphics{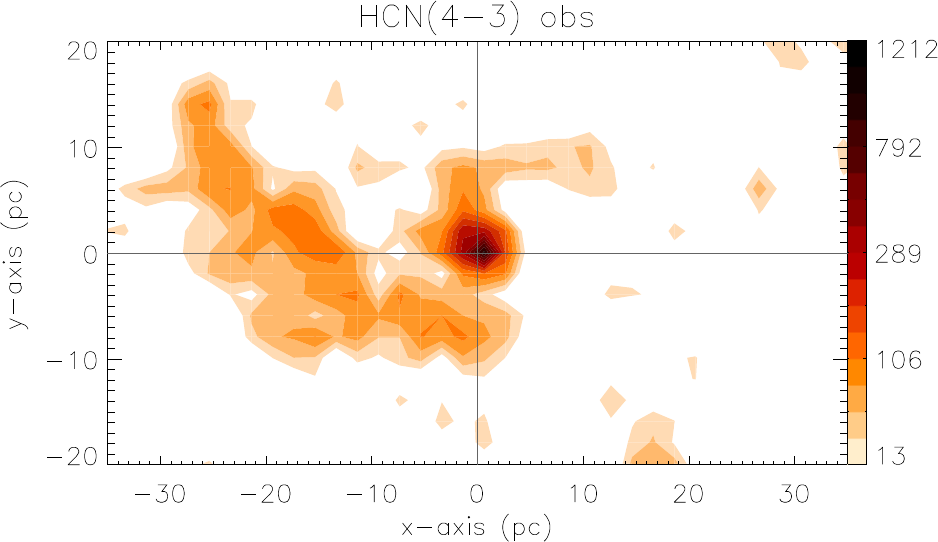}\includegraphics{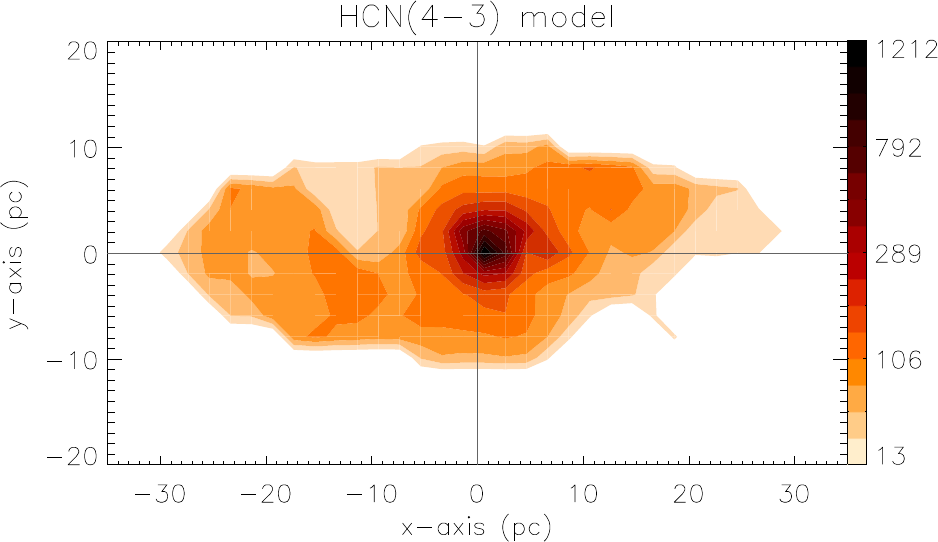}\includegraphics{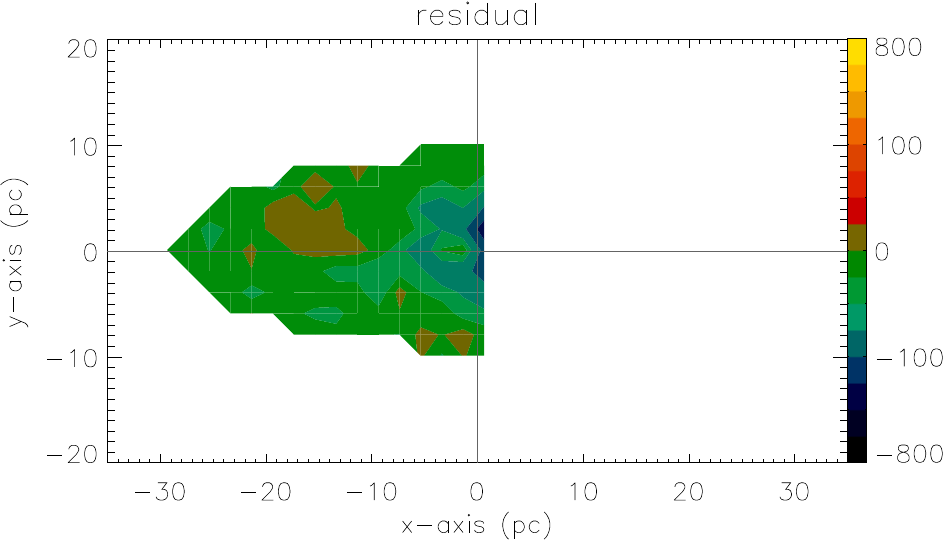}}
  \resizebox{\hsize}{!}{\includegraphics{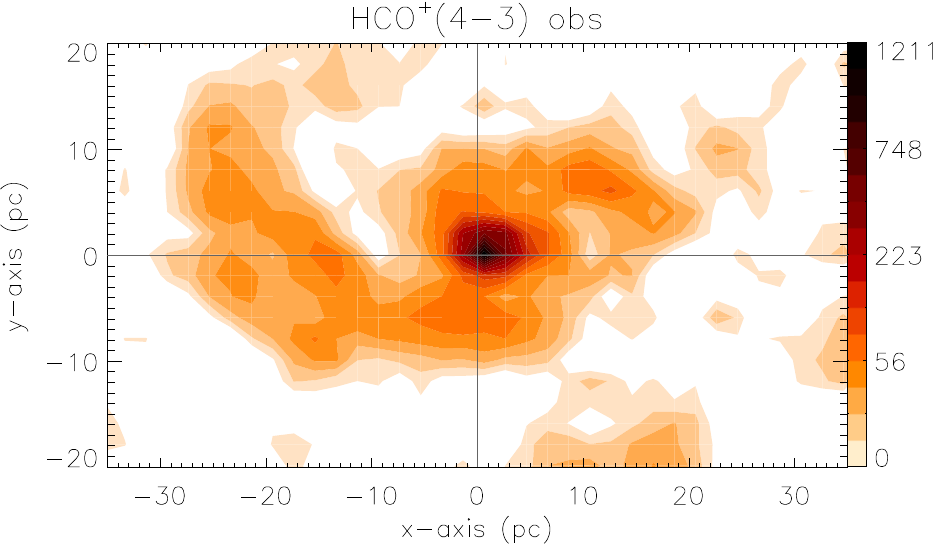}\includegraphics{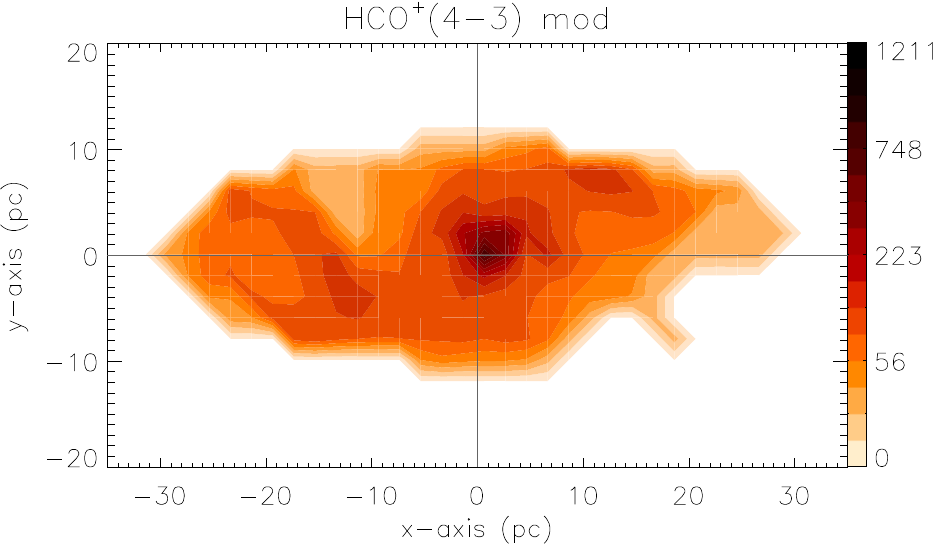}\includegraphics{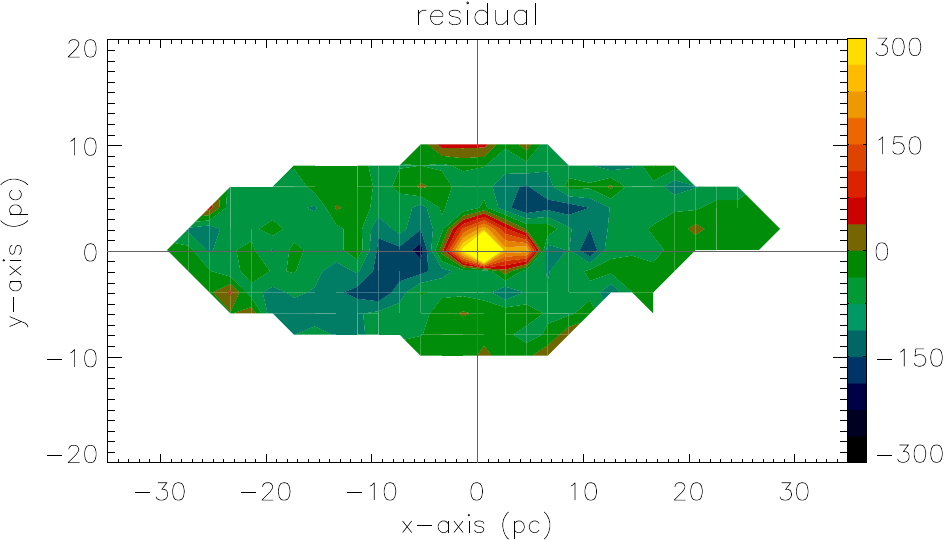}}
  \caption{Comparison between the observed and model maps of molecular line emission.
    The model only applies in the inner $30$~pc shown as black ellipse in the upper left panel.
    Left panels: observations; middle panels: model; right panels: residuals (model$-$observations).
    The units of the color bars are K\,km\,s$^{-1}$.
  \label{fig:fig3}}
\end{figure*}

In a second step, we produced model maps of the mean of $10$ best-fit LR models and the associated scatter (Fig.~\ref{fig:fig3}).
To do so, we rotated the ALMA moment~0 maps by $55^{\circ}$ such that the kinematical major axis is horizontal.
Our comparison is limited to the inner $30$~pc. 

\FloatBarrier

Overall, the observed maps of molecular line emission are well reproduced by the LR model. As already seen in the radial profile of $\sqrt{\chi^2}$
(upper panel of Fig.~\ref{fig:khideux_log_nd_pb6_pco32HR_konrad_nooutflow_2disks_18}),
the inner region with $R \la 4$~pc is less well reproduced by the LR model. Generally, the residuals show a homogeneous scatter except for the
CO(6-5) line where we observe a top-bottom asymmetry. In addition, the northeastern interarm region is not well reproduced in the
CO(3-2) and HCO$^{+}$(3-2) lines, where the model overestimates the flux compared to the observations (yellow and red residuals).

Since the HCN(4-3) line was positioned at the edge of its spectral window, only the blue-shifted part of the line up to the systemic velocity could be covered
(Tristram et al. 2022).
Our model permits to predict the missing data for $v > v_{\rm sys}$ at $x > 0$~pc (HCN(4-3) panel of Fig.~\ref{fig:fig3}).

\begin{figure}
  \centering
  \resizebox{7cm}{!}{\includegraphics{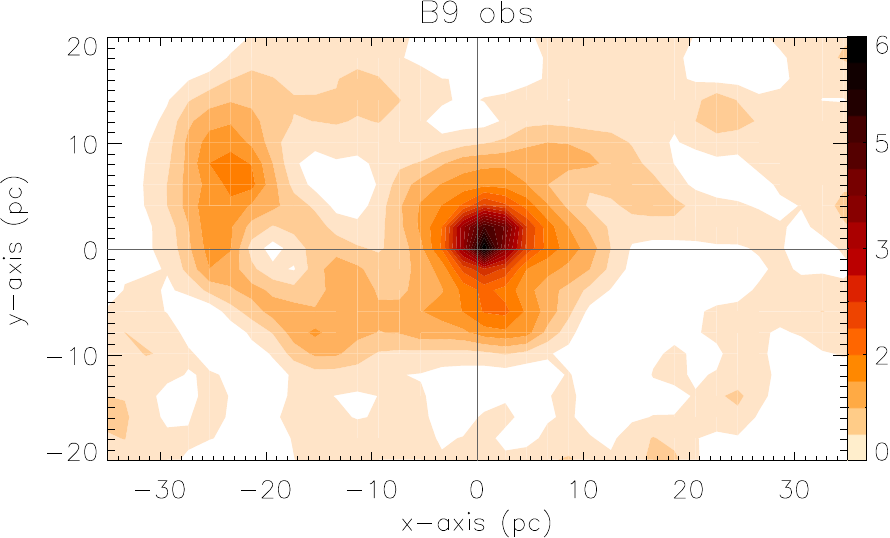}}
  \resizebox{7cm}{!}{\includegraphics{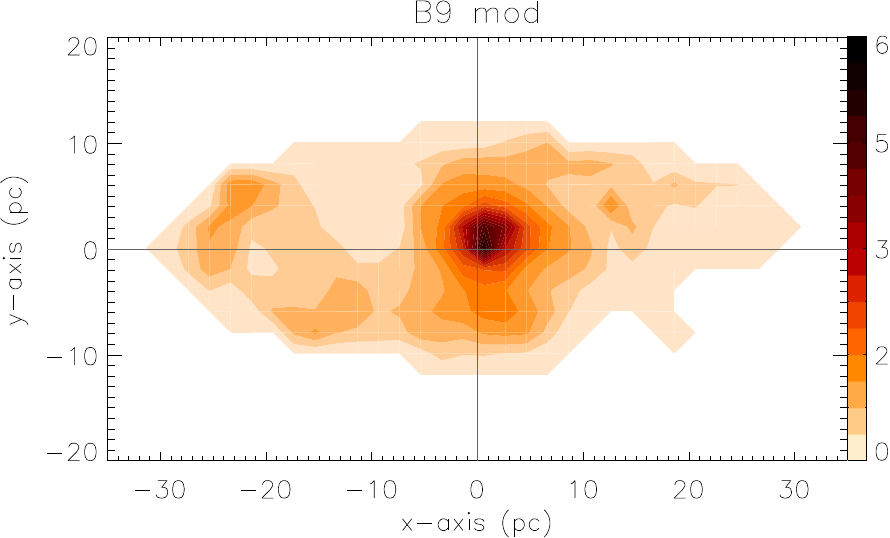}}
  \caption{Comparison between the observed and model maps of band~9 continuum emission. Upper panel: observations; lower panel: model.
    The units of the color bars are K.
  \label{fig:fig6}}
\end{figure}

Since band~9 is the least affected by free-free emission, we show the observed LR band~9 continuum map together with the mean
of the $10$ best-fit models in Fig.~\ref{fig:fig6}. Overall, the observed structure with a strong central emission, a prominent
eastern arm, and a faint western arm is recovered. However, the model eastern arm and central emission are less pronounced than observed.

We conclude that the model can reproduce the observed continuum and molecular line radial profiles within the uncertainties
leading to a total $\sqrt{\chi^2} \la 2$. Generally, The LR model can reproduce the observed molecular line maps leading to
residuals with a homogeneous scatter. The observed HR continuum emission maps are reproduced by the model in a satisfactory way.

\section{Derived parameters \label{sec:parameters}}

\begin{figure*}
  \centering
  \resizebox{\hsize}{!}{\includegraphics{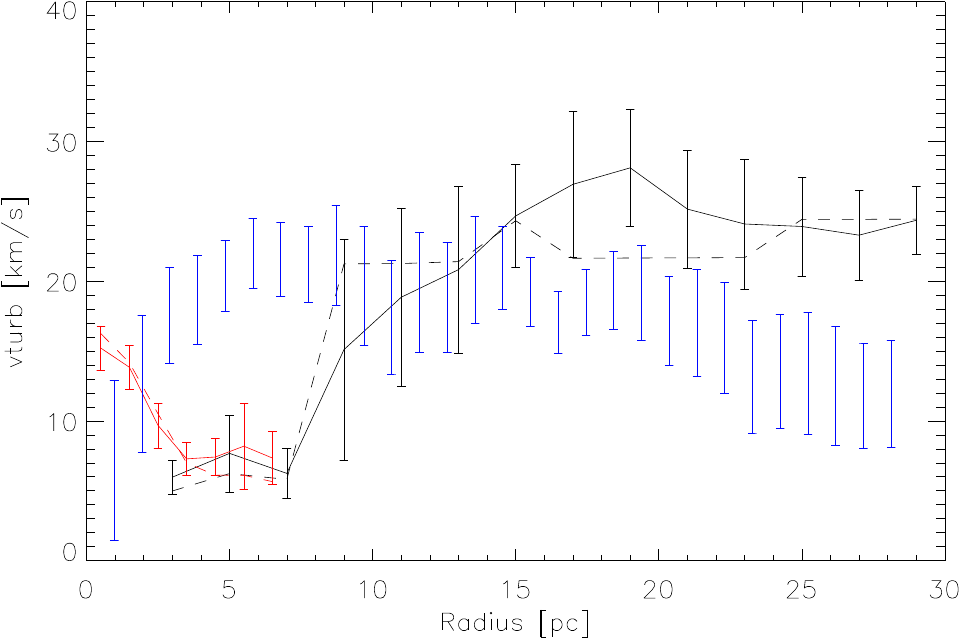}\includegraphics{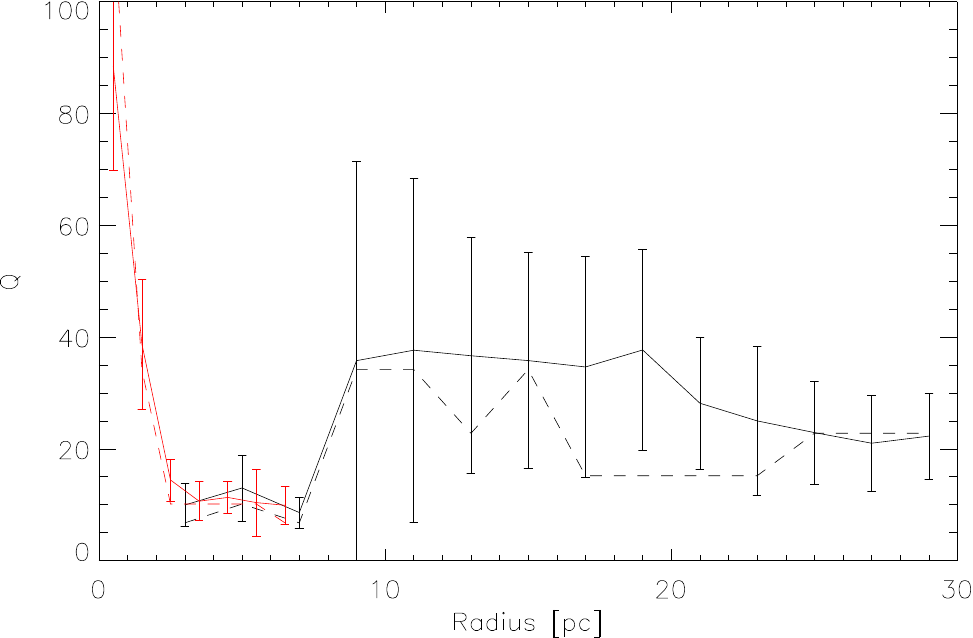}\includegraphics{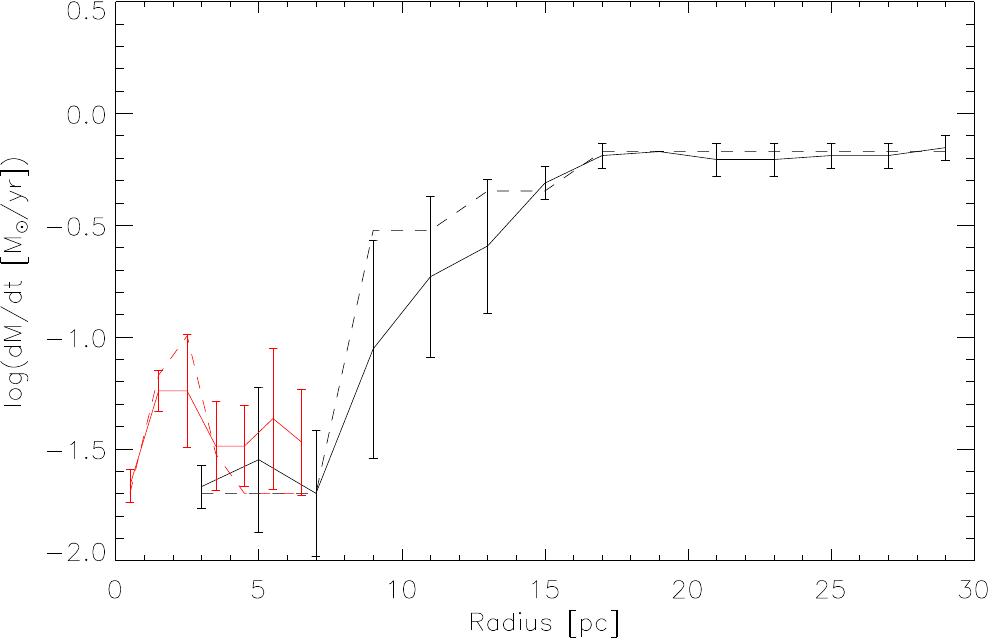}}
  \resizebox{\hsize}{!}{\includegraphics{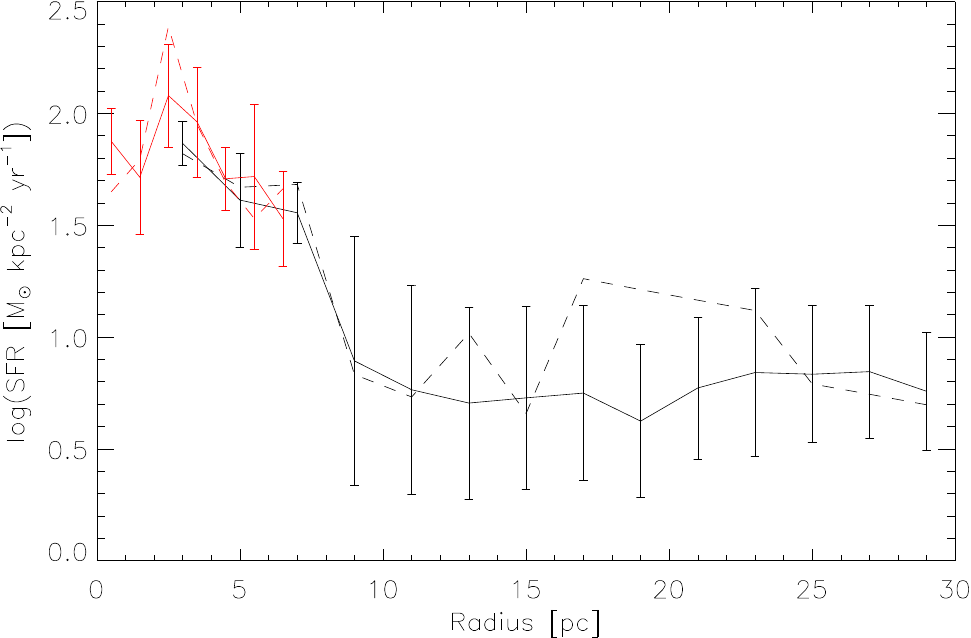}\includegraphics{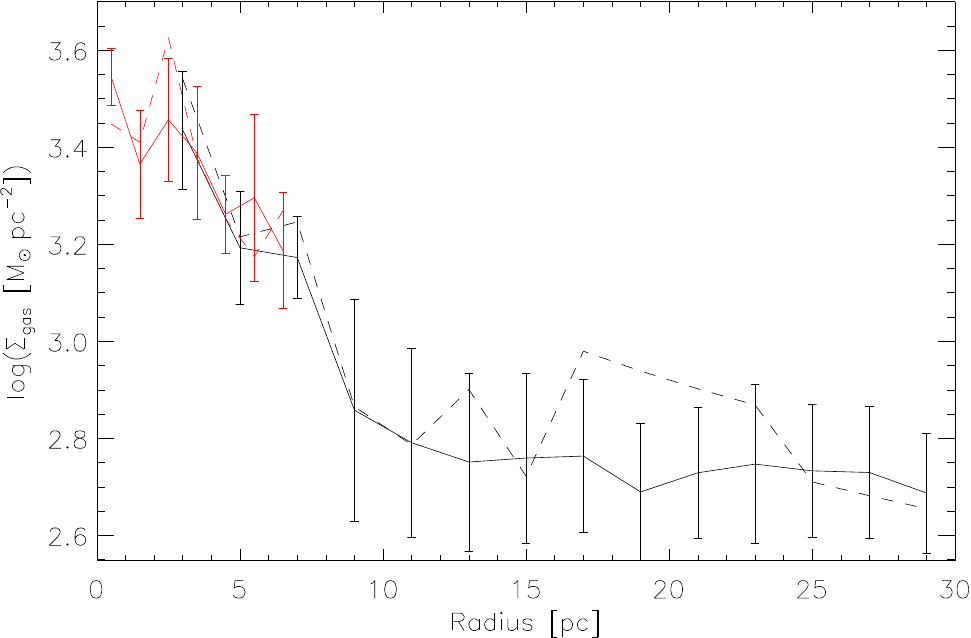}\includegraphics{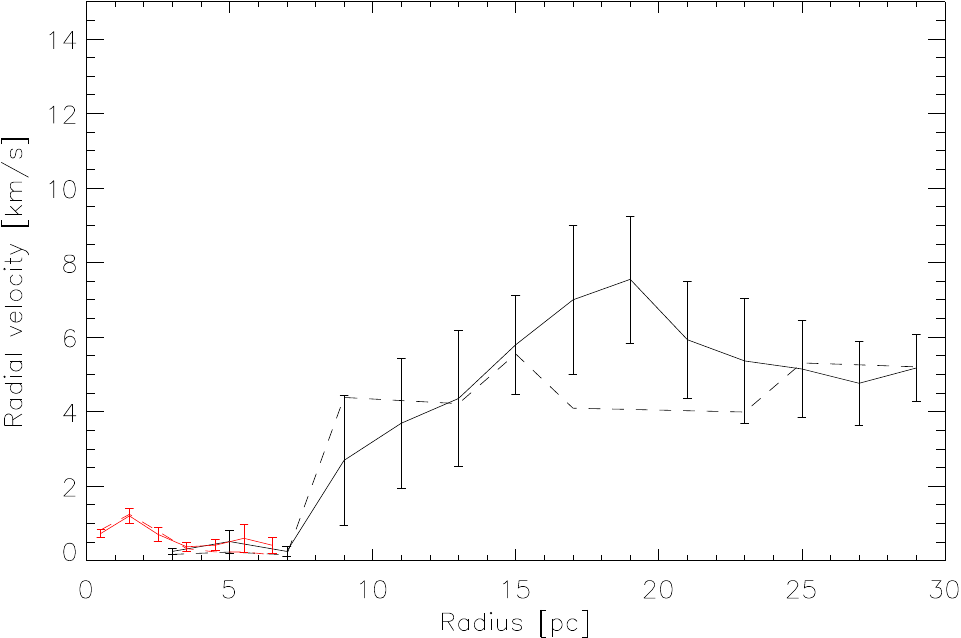}}
  \caption{Radial profiles of the derived physical quantities. From upper left to lower right: gas velocity dispersion (blue error bars: observed HCN(3-2) velocity dispersion from Goesaert et al. 2025), Toomre $Q$ parameter, mass accretion rate $\dot{M}$, star formation rate surface density, gas surface density, gas radial velocity. Red lines: HR model; black lines LR model.
  \label{fig:fig7}}
\end{figure*}

The most important parameters of our model are the Toomre $Q$ parameter and the mass accretion rate $\dot{M}$ within the disk.
With these parameters the gas velocity dispersion $v_{\rm turb}$, gas surface density $\Sigma$, star formation surface density $\dot{\Sigma}_*$,
and gas radial velocity $v_{\rm r}$ can be calculated.
The radial profiles of these parameters are presented in Fig.~\ref{fig:fig7}.

The gas velocity dispersion of the maser disk is $\sim 15$~km\,s$^{-1}$ at $R=0.5$~pc
The velocity dispersion decreases to $8$~km\,s$^{-1}$ at $R=3$~pc. It then stays constant until $R=7$~pc.
After an increase to $\sim 25$~km\,s$^{-1}$ at $R=15$~pc it stays again
approximately constant. The Toomre $Q$ parameter is highest ($Q=80$) in the center, declines to $Q=10$-$15$ at $R=3$~pc, and then stays constant until $R=7$~pc
(top middle panel of Fig.~\ref{fig:fig7}).
After a steep increase it has an about constant value between $Q=30$ and $40$ for $R \ge 10$~pc. The associated scatter is high for this radial range.
The mass accretion rate for $R \le 7$~pc is about constant with $\dot{M} \sim 0.04$~M$_{\odot}$yr$^{-1}$ (top right panel of Fig.~\ref{fig:fig7}).
It rises to somewhat less than $0.6$~M$_{\odot}$yr$^{-1}$ at $R=15$~pc and then stays constant. For a discussion of the increasing
mass accretion rate with increasing radius see Sect.~\ref{sec:inout}.

At radii $3$~pc$ < R < 8$~pc many parameters take lower values compared to regions located further in or out. This corresponds to the
  region where the gravitational potential of the inner disk dominated. The replacement of the second inner stellar disk by a nuclear star cluster
  leads to an increase of these parameters (especially the Toomre $Q$) in this region (see Sect.~\ref{sec:NSC} and
  Fig.~\ref{fig:khideux_log_nd_pb6_pco32HR_konrad_nooutflow_2disks_18a}).

As expected, the radial profiles of the star formation and gas surface densities decrease monotonically with increasing radius up to $R \sim 10$~pc.
These quantities are about constant for $R > 10$~pc ($\dot{\Sigma}_* \sim 6$~M$_{\odot}$yr$^{-1}$kpc$^{-2}$ and $\Sigma \sim 500$~M$_{\odot}$pc$^{-2}$.
The gas radial velocity is about $1$~km\,s$^{-1}$ within the maser disk ($R<1$~pc) and $0.5$~km\,s$^{-1}$ for $2 \le R \le 7$~pc,
increases to $\sim 6$~km\,s$^{-1}$ at $R \sim 17$~pc, and then stays constant.

\begin{figure}
  \centering
  \resizebox{7cm}{!}{\includegraphics{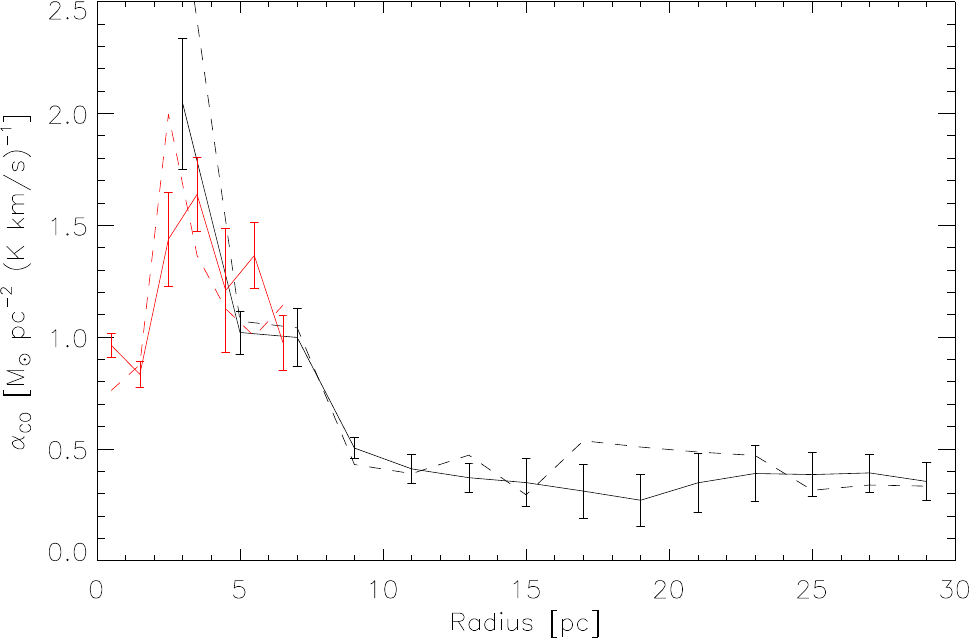}}
  \caption{Radial profile of the model CO(3-2) conversion factor. Red lines: HR model; black lines LR model.
  \label{fig:fig9}}
\end{figure}

Since the model gas surface densities and CO fluxes are known, the model CO(3-2) conversion factor can be calculated (Fig.~\ref{fig:fig9}).
The model CO(3-2) conversion factor is $\alpha_{\rm CO} \sim 0.5$~M$_{\odot}$pc$^{-2}$/(K\,km\,s$^{-1}$) for $R \ge 9$~pc. It increases with decreasing radius to
$1.5$~M$_{\odot}$pc$^{-2}$/(K\,km\,s$^{-1}$). The conversion factor is $\alpha_{\rm CO} \sim 1$~M$_{\odot}$pc$^{-2}$/(K\,km\,s$^{-1}$) within the maser disk.

\begin{figure*}
  \centering
  \resizebox{14cm}{!}{\includegraphics{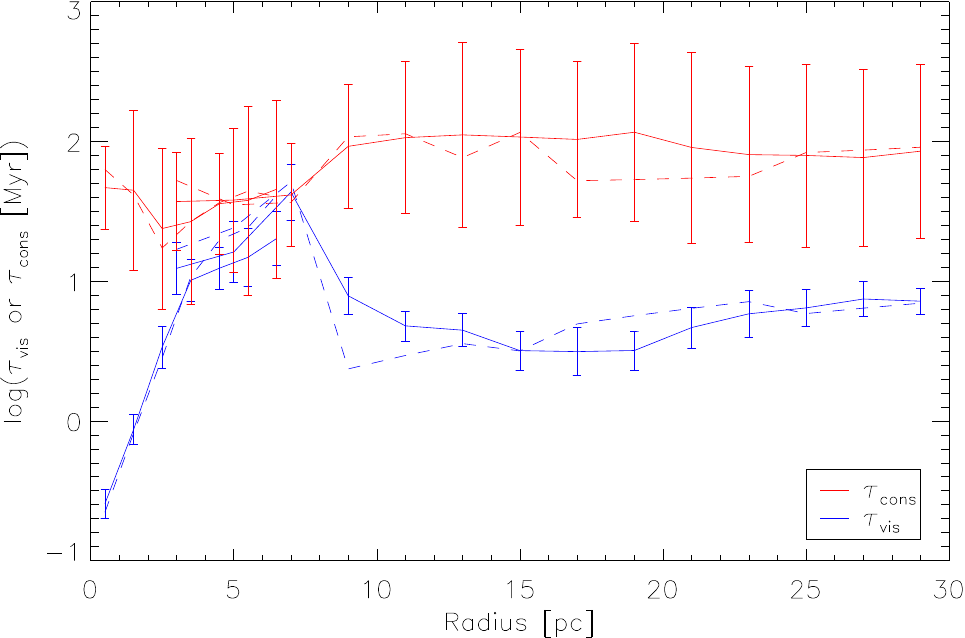}\includegraphics{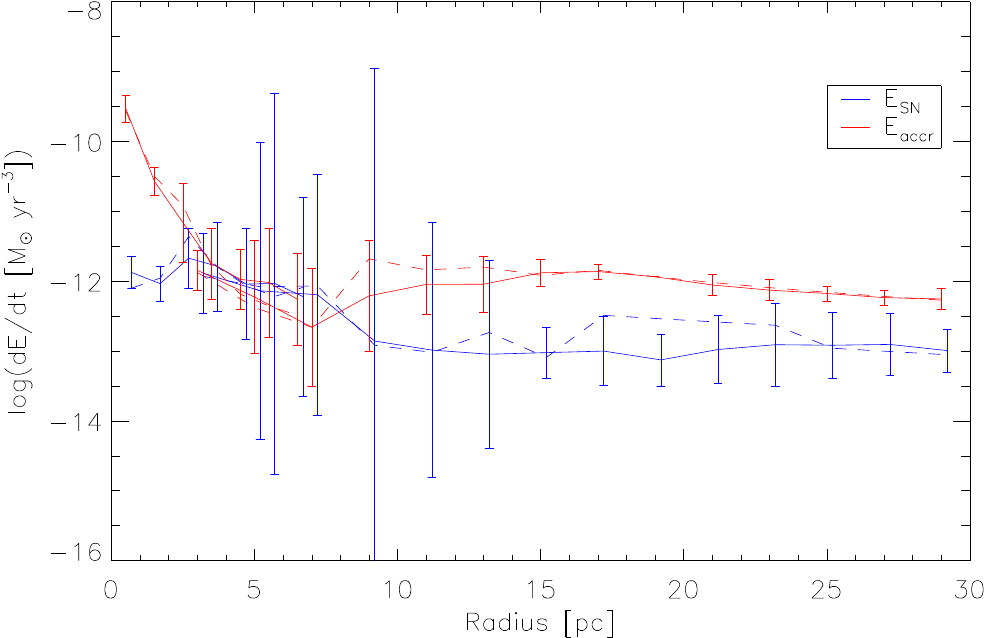}}
  \caption{Radial profiles of the model viscous and gas consumption timescales (left panel) and energy injection rates via SN and accretion (right panel).
  \label{fig:fig10}}
\end{figure*}

In massive star-forming galactic disks the gas consumption timescale $t_*$ is much smaller than the viscous timescale $t_{\rm visc}$ (Liz\'ee et al. 2022).
Thus, radial gas transport cannot compensate for gas consumption via star formation.
To investigate if this is also true in the central $30$~pc of the Circinus galaxy, we calculated these timescales in the following way
\begin{equation}
  t_{\rm visc}=\frac{(R^2 \Omega)'}{-\Omega'} \frac{1}{\nu}\ ,\ \ \ 
  t_*=\frac{\Sigma}{\dot{\Sigma}_*}\ ,
\end{equation}
where $\Omega=v_{\rm rot}/R$ is the angular velocity, and $\nu=v_{\rm turb}l_{\rm driv}$ the gas viscosity.
The resulting timescales are presented in the left panel of Fig.~\ref{fig:fig10}.
For radii $R > 7$~pc the viscous timescale is significantly smaller than the gas consumption timescale.
Between $3$ and $7$~pc both timescales are about equal. The viscous timescale is again smaller than the gas consumption time scale in the
maser disk ($R \le 1$~pc).

The viscous timescale ($t_{\rm visc} \la 1$~Myr) is much longer than the dynamical timescale ($t_{\rm dyn} \sim 0.1$~Myr).
This implies that the CND is in a quasi-steady state for about $10$ dynamical times. Since the typical lifetime of stars that will explode as 
SN type II is a few Myr, strong variations of the star formation rate on smaller timescales cannot be modelled by our approach.

The radial profiles of the model metallicity normalization $k_{\rm Z}$ and CR injection rate normalization $k_{\rm CR}$ are presented in Figs.~\ref{fig:fig10a} and \ref{fig:fig10b}.
Both quantities show quite large scatters. The metallicity normalization is about constant at $k_{\rm Z} \sim 2$.
The CR injection rate with respect to the SF volume density is about $1.5$ to $2$ times higher than expected from  the Vollmer et al. (2025) model for $R \le 7$~pc and
about three times smaller than expected for $R > 10$~pc.

The energy injection rate via SN explosions is given by Eq.~\ref{eq:energyflux}. That by radial accretion and subsequent gain of
gravitational potential energy is given by
\begin{equation}
  \frac{\dot{E}_{\rm acc}}{\Delta A}=\frac{1}{4}\pi \dot{M} (R \Omega')^2\ .
\end{equation}
For radii $R > 7$~pc and within the maser disk ($R < 1$~pc) the gain of gravitational energy via accretion dominates the energy injection rate
(right panel of Fig.~\ref{fig:fig10}).
For $2 \le R \le 7$~pc both energy injection rates are about the same.

\begin{figure*}
  \sidecaption
    \begin{minipage}{12cm}
    \includegraphics[width=6cm,height=!]{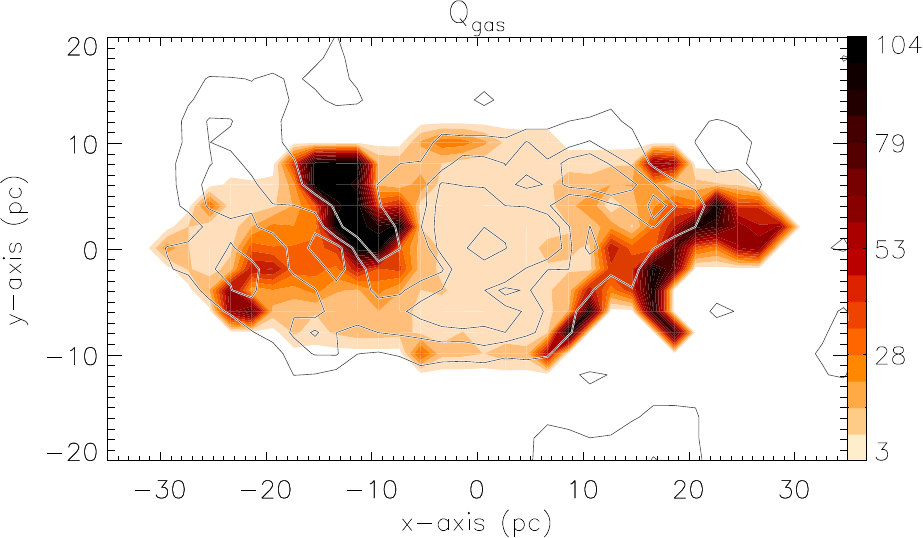}
    \hspace{2mm}
    \includegraphics[width=6cm,height=!]{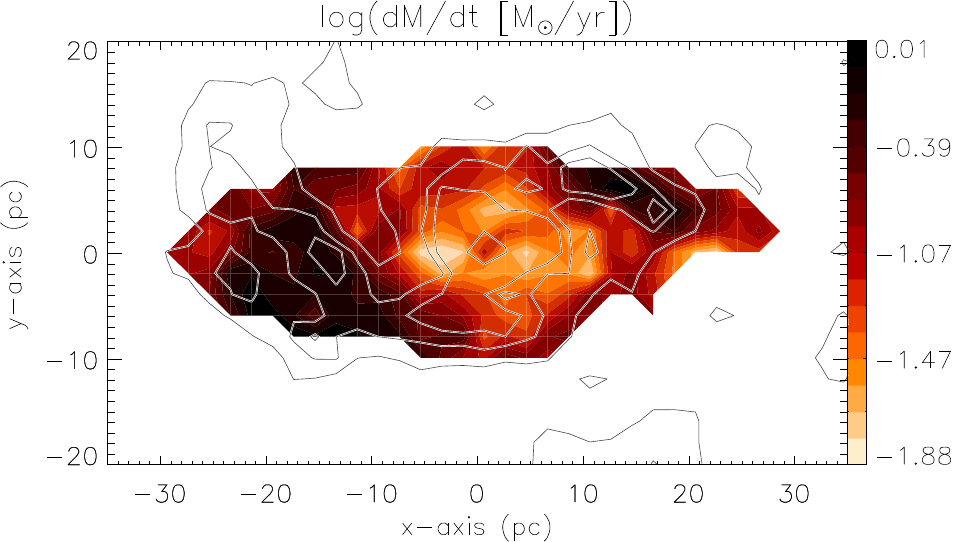}
    \vspace{2mm}
    \includegraphics[width=6cm,height=!]{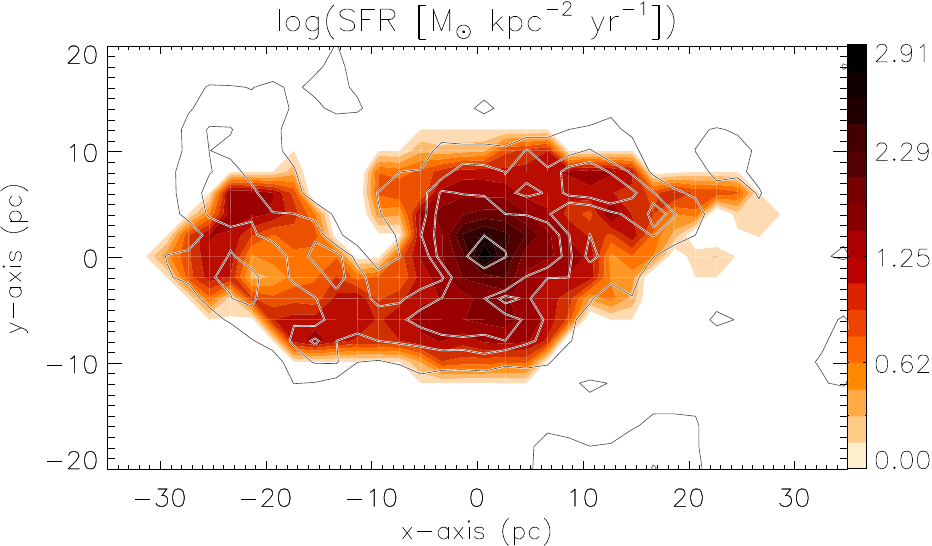}
    \hspace{2mm}
    \includegraphics[width=6cm,height=!]{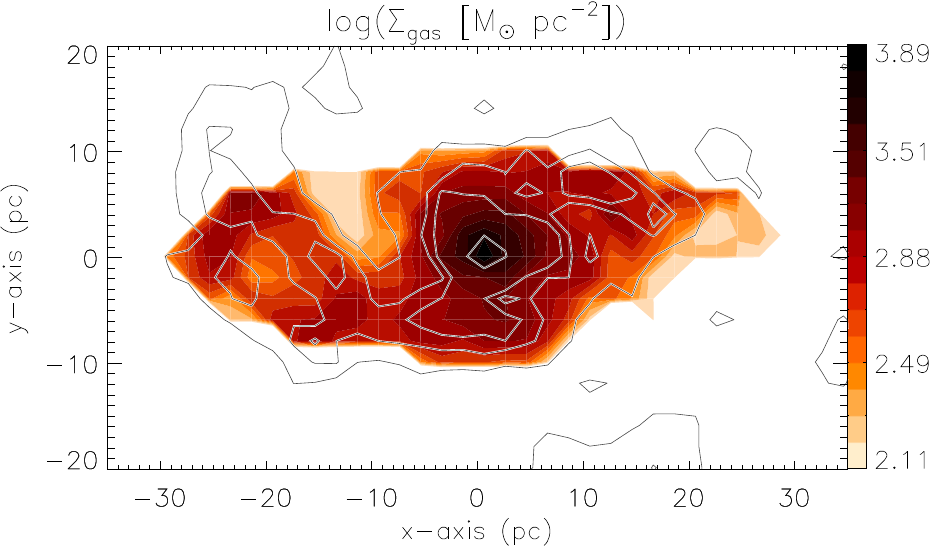}
    \vspace{2mm}
    \includegraphics[width=6cm,height=!]{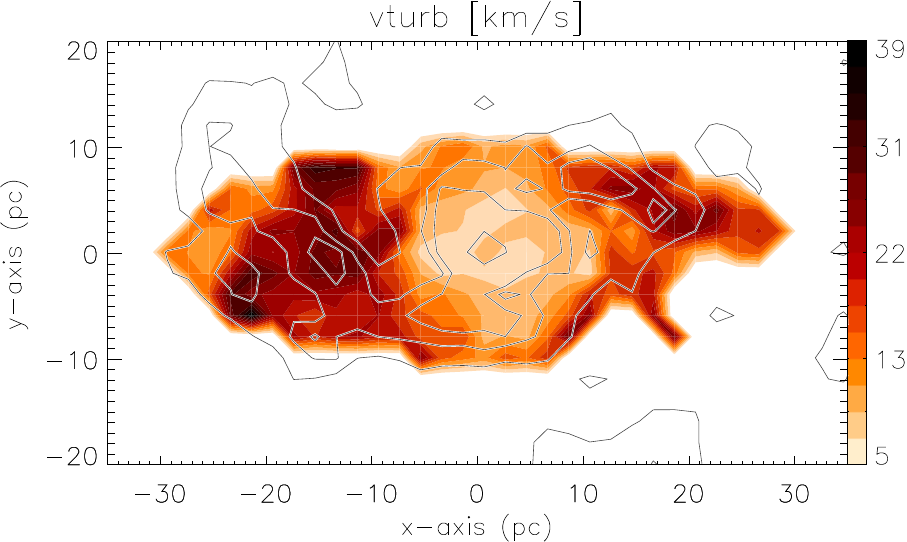}
    \hspace{2mm}
    \includegraphics[width=6cm,height=!]{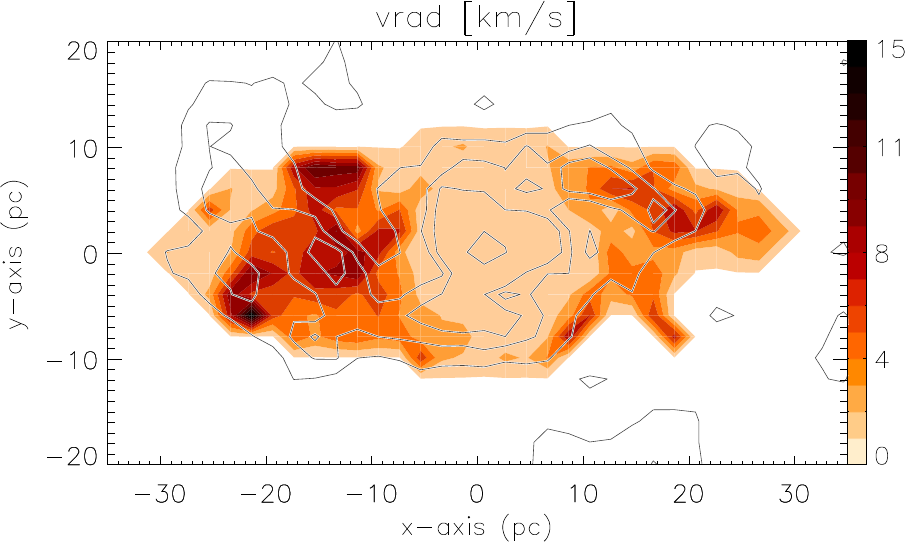}
    \vspace{2mm}
    \includegraphics[width=6cm,height=!]{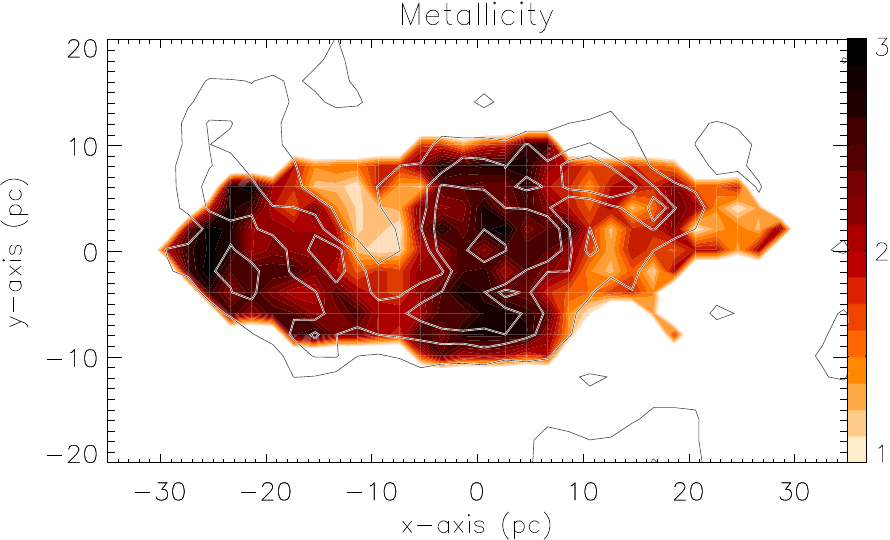}
    \hspace{2mm}
    \includegraphics[width=6cm,height=!]{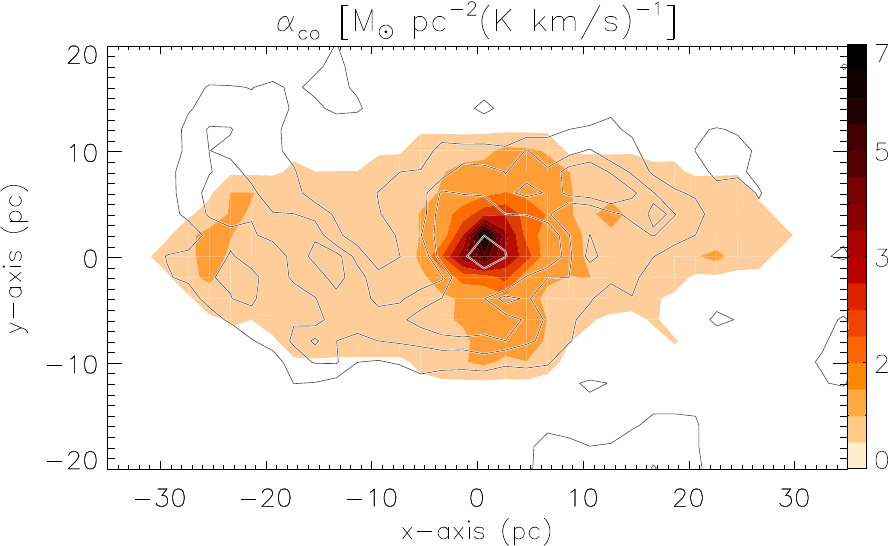}
  \end{minipage}
  \caption{Maps of the derived physical quantities. The HCO$^+$ (4-3) emission is shown by contours to guide the eye.
    The units of the color bars are the same as stated in the titles of the panels.
  \label{fig:fig11}}
\end{figure*}

The map of the model CR injection normalization is presented in Fig.~\ref{fig:fig10b}.
  The CR injection normalization is highest in the center and in the inner parts of the two gas arms.
   The metallicity is lowest in the interarm regions and the
CO(3-2) conversion factor is highest in the center and in the inner parts of the two gas arms.
The maps of the derived physical quantities are shown in Fig.~\ref{fig:fig11}.
As expected, the star formation rate and gas surface densities are
highest in the center and in the two gas arms.
The Toomre $Q$ parameter is lowest in the center and the northern gas arm, whereas it is highest in the northeastern interarm region.
The gas turbulent and radial velocities are highest in the outer parts of the two gas arms.
This is expected because both are closely linked to the mass accretion rate.
Consequently, the mass accretion rate $\dot{M}$ is highest in the outer parts of the two gas arms.

Overall, the model shows that the CND is inhomogeneous, with gas inflow, turbulence, and star formation concentrated in the gas arms and the
  central region, whereas interarm regions are relatively quiescent.

\section{Discussion \label{sec:discussion}}

In the following we give an overview on the result of the model including a nuclear star cluster and present and discuss the properties of the ISM
at $R \le 30$~pc. The star formation rate derived from different tracers is compared to our model results. Moreover, the exponent of the
Schmidt Kennicutt law is derived and we hint at the possibility of a SN-driven galactic wind. Finally, the gas inflow and outflow near the dust sublimation
radius are discussed and we comment on the obscuration of the AGN.

\subsection{Nuclear star cluster model \label{sec:NSC}}

We wanted to test, if a more Milky Way like stellar distribution can also explain the data in the Circinus Galaxy.
The stellar distribution of the Galaxy is comprised of a nuclear stellar disk
  ($R_0=74$~pc and $\Sigma_*=1.8 \times 10^4$~M$_{\odot}$pc$^{-2}$, Sormani et al. 2022)
and a nuclear star cluster ($M_*=2 \times 10^7$~M$_{\odot}$ and half light radius of $4$~pc;  Sch\"odel et al. 2014).
Therefore, we replaced the second stellar disk of the fiducial model by a nuclear star cluster that is consistent with the observed rotation curve.
It has a stellar density of the following form
\begin{equation}
  \rho_*(R)=\rho_{*,0}\,\exp{\big(-\frac{2}{3}\,(R/R_0)^{1/3}-1\big)}
\end{equation}
with $\rho_{*,0}=5\times 10^3$~M$_{\odot}$pc$^{-3}$ and $R_0=4$~pc. To reproduce the observed rotation curve at $R \ga 10$~pc,
we multiplied the surface density of the first nuclear disk by $50$\,\% with respect to that of the fiducial model.
Our nuclear star cluster has a mass of $M_*=10^7$~M$_{\odot}$ and a half-mass radius of $12$~pc, which is comparable to the effective radius
of $8$~pc determined by Pechetti et al. (2020). The central density was chosen such that it is consistent with the rotation curve.

The derived physical quantities for this model are presented in Fig.~\ref{fig:fig7a}.
Since the nuclear star cluster has a much higher velocity dispersion than the second nuclear stellar disk of the fiducial model and thus is vertically more
extended than the second nuclear disk, this model has a smaller vertical acceleration of the gas disk compared to the fiducial
model in the inner $\sim 10$~pc. The resulting gas velocity dispersions are higher for $5 \le R \le 10$~pc and lower for $R >10$~pc than those of the fiducial model.
The differences at large radii are due to the higher mass surface density of the nuclear stellar disk with respect to that of the fiducial model.
The resulting Toomre $Q$ parameters are about a factor of two higher for $2 \le R \le 7$~pc and by the same factor lower for $10 \le R \le 20$~pc than those of the fiducial model.
The resulting mass accretion rate at $R > 15$~pc is about a factor of about three smaller than that of the fiducial model. The associated error bars are much larger than
and include the mass accretion rates of the fiducial model. The resulting CR ionization rate is about that expected by the Vollmer et al. (2025) model.

Whereas the gas mass surface densities of both models are comparable for $R \ga 10$~pc, the star formation surface density of the model including a nuclear star cluster is about
a factor of two higher than that of the fiducial model. The radial velocities are about a factor of three smaller for $R \ga 15$~pc than those of the fiducial model.
The resulting viscous timescale is close to but smaller than the gas consumption timescale. The resulting energy injection rates by SN and gain of gravitational energy through
accretion are about the same. Especially the SN energy injection rate has a huge scatter, i.e. it is not well constraint.

We conclude that overall both models yield consistent results within a factor of about two.
Therefore, the detailed form of the stellar potential does not affect the main conclusions of the model.

\subsection{ISM properties \label{sec:ISM}}

The radial profile of the gas velocity dispersion $v_{\rm turb}$ in the inner $30$~pc of Circinus has been studied by Tristram et al. (2022), Izumi et al.
(2023), and Goesaert et al. (2025). Tristram et al. (2022) reported values of $10$ to $25$~km\,s$^{-1}$ depending on the molecular tracer.
Using HCN(3–2) and HCO$^+$(3–2), Goesaert et al. (2025) find dispersions of $\sim 20$~km\,s$^{-1}$ between $3$ and $20$~pc, decreasing to
$\sim 10$~km\,s$^{-1}$ in the maser disk ($R \sim 1$~pc; upper left panels of Figs.~\ref{fig:fig7} and \ref{fig:fig7a}). Izumi et al. (2023) also derive a value of $\sim 10$~km\,s$^{-1}$ for the maser disk based on
HCN(3–2). Our model reproduces a similar overall radial trend, although the predicted velocity dispersion is about 50\,\% higher than that of
Izumi et al. (2023) and Goesaert et al. (2025) in the maser disk ($R<1$~pc), and about a factor of two lower than the value of Goesaert et al. (2025) at
$3$~pc$ < R < 7$~pc. At larger radii, our values are slightly higher than those of Goesaert et al. (2025). Overall, the model agrees with the observational
estimates within $\sim 50$\,\%, except around $R \sim 5$~pc where the discrepancy reaches about a factor of two. At this radius Goesaert et al. (2025)
found important non-circular motions (their Fig.~8). Their derived velocity dispersion is thus most probably caused by these non-circular motions.
Goesaert et al. (2025) interpreted these motions as inflow. In this case, our mass accretion rate should be taken as a lower limit.

Izumi et al. (2018) estimated the total mass in the CND by assuming that the CO(3–2) emission is thermalized with that of CO(1–0), using the standard equation (Solomon
\& Vanden Bout 2005) to calculate the line luminosity, and applying the canonical CO conversion factor in active environments
($\alpha_{\rm CO} \sim 0.8$~M$_{\odot}$pc$^{-2}$/(K\,km\,s$^{-1}$); Downes \& Solomon 1998, Bolatto et al. 2013). Indeed, we found a CO conversion factor
of $\alpha_{\rm CO} \sim 0.5$~M$_{\odot}$pc$^{-2}$/(K\,km\,s$^{-1}$), which is close to the canonical CO conversion factor in starburst environments.
We thus confirm the assumption of Izumi et al. (2018). With our conversion factor and the CO(3-2) flux determined by Izumi et al. (2018) the total gas mass of the CND is 
$M_{\rm gas} \sim 2.7 \times 10^6$~M$_{\odot}$. The total stellar mass within $30$~pc is $M_*=8.4 \times 10^7$~M$_{\odot}$ and the gas fraction
$M_{\rm gas}/M_*=0.03$. We verified that the gas is not vertically selfgravitating within $R<30$~pc. This means that the stellar gravitational potential
dominates the vertical acceleration of the gas everywhere in the disk.

Whereas the critical density of CO(3-2) in the optically thin limit\footnote{In practice the critical density is decreased to a few
    $10^3$~cm$^{-3}$ by line trapping.} is about $3 \times 10^4$~cm$^{-3}$, that of HCO$^+$(3-2) is about $2 \times 10^6$~cm$^{-3}$.
The HCO$^+$(3-2) line emission thus traces much denser gas than the CO(3-2) line emission. We investigated the correlation between these two lines and the physical
quantities of the disk. CO(3-2) is not strongly correlated with the gas mass surface density nor with the star formation surface density.
In the following, we only report correlations with a Pearson correlation coefficient higher than $0.6$.
The HCO$^+$(3-2) emission is strongly correlated with the gas mass surface density, the star formation surface density, and gas pressure. 
It is strongly anticorrelated with the gas consumption timescale. The physical properties of the galactic gas disk at $R \le 30$~pc are thus best traced by
the line emission of high critical density.

We conclude that our model gas velocity dispersion, CO conversion factor, and gas mass of the CND are consistent with results from the literature.
The HCO$^+$(3-2) line emission traces best the ISM properties.

\subsection{Star formation \label{sec:SFR}}

We found a star formation rate for $R \le 30$~pc of $\dot{M}_*=0.03$~M$_{\odot}$yr$^{-1}$. This is comparable to the star formation rate estimated by
Maiolino et al. (2000) based on the Pa$\alpha$ emission $\dot{M}_* \sim 0.01$~M$_{\odot}$yr$^{-1}$. Their value is lower because they assumed a lower
gas mass surface density compared to our model results leading to an underestimation of the extinction correction.
The mean star formation rate within $R=8$~pc of $\dot{M}_*=0.014$~M$_{\odot}$yr$^{-1}$ estimated by M\"{u}ller Sanchez et al. (2006) and Davies et al. (2007)
is also consistent with our model star formation rate. This estimate was based on the Br$\gamma$ equivalent width and a SN rate of $\nu_{\rm SN}=2.4 \times 10^{-4}$~yr$^{-1}$.

\subsubsection{Radio continuum emission}

In the following we investigate if radio continuum emission can be used as a tracer for star formation in the inner $R \le 10$~pc of the Circinus galaxy
by comparing the relevant timescales. We normalize the Circinus SN rate by the Galactic value.
The integrated rate of SNe type II in the Galaxy is taken to be $\nu_{\rm SN} \sim 1/60$~yr$^{-1}$ (Rozwadowska et al. 2021).
The Galactic SFR is taken to be $\dot{M}_* = 2$~M$_{\odot}$yr$^{-1}$ (Licquia \& Newman 2015, Elia et al. 2022).
This leads to a characteristic time between two subsequent SN explosions for a star formation rate of $\dot{M}_* \sim 0.01$~M$_{\odot}$yr$^{-1}$ within $R=10$~pc in
Circinus of $t_{\rm SN} \sim 10^4$~yr. The SN fading time (Eq.~39.30 of Draine et al. 2011) when the declining shock velocity reaches the turbulent velocity
is $t_{\rm SN,fade} \ga 6 \times 10^4$~yr.

Assuming equipartition
between the turbulent kinetic and magnetic energy densities, the magnetic field strength is $B \sim 1$~mG. This leads to a synchrotron timescale at $8.6$~GHz of
$t_ {\rm sync} \sim  10^4$~yr. With a gas density of $n_{\rm H} \sim 10^4$~cm$^{-3}$ the bremsstrahlung timescale is even shorter $t_{\rm brems} \sim 4 \times 10^3$~yr.
The distribution of the gas surface density is regulated by the viscous timescale ($t_{\rm visc} \sim 8$~Myr; upper left panel of Fig.~\ref{fig:fig10}),
which is much longer than the dynamical timescale ($t_{\rm dyn} \sim 0.1$~Myr). The gas depletion timescale is
about twice the viscous timescale. The lifetime of an ionizing star is $t_{\rm star} \sim {\rm a\ few}$~Myr before it explodes as a SN.
Since $t_{\rm visc} > t_{\rm star}$, SN explosions can be used as a tracer of star formation.
Since $t_{\rm SN} > min(t_{\rm sync},t_{\rm brems})=t_{\rm brems}$, the nonthermal radio continuum emission at $8.6$~GHz is not a good tracer of the star formation rate.
However, at these high gas densities the radio continuum emission at $8.6$~GHz is expected to be dominated by thermal free-free emission.
Thus, only if the radio continuum emission is dominated by free-free emission, it is a good tracer of star formation. 

In the following we will investigate the nature of the radio continuum emission.
M\"{u}ller Sanchez et al. (2006) reported a $8.6$~GHz continuum luminosity of $L_{8.6}=2.5 \times 10^{19}$~W\,Hz$^{-1}$
for $R \le 8$~pc derived from the radio continuum data of Davies et al. (1998).
We calculated the total radio continuum emission based on the model of Vollmer et al. (2025) using the our HR model radial profiles.
The radio continuum model takes into account the magnetic field strength, CRe injection rate, and CRe energy loss terms.
The model yielded a total radio continuum luminosity of $L_{\rm 10,model}=2.4 \times 10^{18}$~W\,Hz$^{-1}$ with a thermal fraction of $70$\,\%
and a spectral index of $0.1$ between $5$ and $10$GHz. Our model star formation rate is $\dot{M}_* = 0.01$~M$_{\odot}$yr$^{-1}$ within $R \le 8$~pc.
The observed radio continuum luminosity is a factor of ten higher than that obtained by the model.
We thus conclude that the $8.6$~GHz radio continuum emission at this spatial scale is dominated by thermal or nonthermal emission linked to the AGN
and not to star formation.
This emission is either optically thin free-free emission or optically thick synchrotron emission leading to a flat spectral index (GHz-peaked radio source).

\subsubsection{Star formation efficiency}

Vollmer et al. (2025) showed that the $\log \dot{\Sigma}_*$--$\log \Sigma_{\rm H_2}$ relation cannot be ﬁt
with a single slope or offset. The slopes of the relation are
$1.2$, $1.7$, $2.0$, $1.1$, $1.4$, and $2.3$ for local star-forming galaxies and starbursts,
local and z$\sim 0.5$ luminous infrared galaxies (LIRGs), and high-z star-forming galaxies and starbursts,
respectively. The local and high-z starbursts and the local LIRGs show slopes  $\sim 2$.
Vollmer et al. (2025) interpreted these steep slopes as a result of the compactness of these galaxies.

The $\log \dot{\Sigma}_*$--$\log \Sigma_{\rm H_2}$ relation in the inner $30$~pc of the Circinus galaxy is presented in Fig.~\ref{fig:fig12}.
Its slope, as derived by the IDL routine robust\_lineﬁt, is $1.9$. The relation is as steep as those of local and high-z starbursts and the local LIRGs.
This is consistent with the low CO conversion factor (see Sect.~\ref{sec:ISM}), which also applies to this kind of compact and active environment.
Despite the high Toomre parameter of $\sim 30$, the mean star formation rate and gas mass surface densities of $\dot{\Sigma}_*=10$~M$_{\odot}$yr$^{-1}$kpc$^{-2}$
and $\Sigma=800$~M$_{\odot}$pc$^{-2}$ are very high placing the inner $30$~pc of the Circinus galaxy at the interface between luminous and ultraluminous
infrared galaxies (Fig.~19 of Vollmer et al. 2025). The inner $30$~pc of the Circinus galaxy can thus be qualified as a starburst region.
The derived high value of the Toomre parameter is not in contradiction with the starburst classification because the star formation surface density
also depends on $\Sigma_*$ and $\Omega$, which are both very high in this region (see Eq.~\ref{eq:sfr11}).

The mean star formation surface density is significantly higher than the Heckman limit ($\dot{\Sigma}_*=0.1$~M$_{\odot}$yr$^{-1}$kpc$^{-2}$; Heckman 2002)
and the disk is expected to develop a galactic wind. Such a wind is not detected in the high-resolution ALMA data (Tristram et al. 2022).
It might be possible that at least a part of the large-scale outflow found by Zschaechner et al. (2016) is caused by the galactic wind expelled from the central
region of the Circinus galaxy ($R \la 80$~pc).
Based on the mean star formation surface density, the estimated star formation rate of this central region is $0.2$~M$_{\odot}$yr$^{-1}$.
This is comparable to the lower limit of the molecular outflow rate given by (Zschaechner et al. 2016). The mass load of such a galactic wind would then be $> 1$.

\subsubsection{The maser disk}

The maser disk is found at smaller distance from the black hole. It extends to about $R \sim 0.5$~pc (Greenhill et al. 2003). 
We derive a total gas mass of $M_{\rm gas} \sim 3 \times 10^3$~M$_{\odot}$ and a star formation rate of $\dot{M}_* \sim 6 \times 10^{-5}$~M$_{\odot}$yr$^{-1}$
for $R \le 0.5$~pc and $M_{\rm gas} \sim 1.6 \times 10^4$~M$_{\odot}$ and a star formation rate of $\dot{M}_* \sim 3.6 \times 10^{-4}$~M$_{\odot}$yr$^{-1}$
for $R \le 1.5$~pc.
Its velocity dispersion is $\sim 15$~km\,s$^{-1}$ (lower left panel of Fig.~\ref{fig:fig7}).
With rotation velocities $\ge 80$~km\,s$^{-1}$ (Fig.~\ref{fig:rotcurve}) it is thus geometrically thin (see also Izumi et al. 2023).
Our model disk at $R \la 1$~pc has a velocity dispersion of $\sim 10$-$15$~km\,s$^{-1}$. It is thus also geometrically thin.
Following Izumi et al. (2023), we identify this model disk as the maser disk.
It is surprising to us that the Circinus maser disk can be modeled as a turbulent clumpy star-forming accretion disk.
With a Toomre parameter of $Q > 30$ (lower middle panel of Fig.~\ref{fig:fig7}), it has a relatively small gas mass and star formation efficiency.
Nevertheless, it actively forms stars up to distance of $\sim 1$~pc to the central black hole.

\subsubsection{Conclusions on the star formation}

We conclude that the radio continuum emission at $R \le 8$~pc is linked to the AGN and not to star formation.
Our model star formation rate of the CND is consistent with results from the literature. The CND can be qualified as a starburst environment
($\dot{\Sigma}_* \propto \Sigma^2$). 
Rather surprisingly, the maser disk at $R \le 1$~pc can be modeled as a turbulent clumpy star-forming accretion disk.

\subsection{Gas inflow and outflow \label{sec:inout}}

In the large-scale disks of massive spiral galaxies the energy injection rate is dominated by that of exploding supernovae and radial gas
transport does by far not compensate the gas consumption via star formation (Liz\'ee et al. 2022). The situation entirely changes
in the inner $30$~pc of the Circinus galaxy: the energy injection rate is dominated by the gain of gravitational potential energy through
gas accretion within the galactic disk and the viscous timescale is significantly smaller than the gas consumption timescale.
This means that star formation does not significantly alter radial gas accretion. The distribution of the gas mass within the
galactic disk and its velocity dispersion are set by the radial gas accretion. The radial profile of the mass accretion rate in the inner $30$~pc 
is set by the variation of the mass accretion rate at $R=30$~pc (regulated by the nuclear spirals; Izumi et al. 2018),
the radial variation of the viscous timescale in the inner $30$~pc, and
a putative outflow, which can be caused by supernova explosions (e.g. Wada 2001), radiation pressure (Wada et al. 2016), or magnetocentrifugal
forces (Vollmer et al. 2018).

The derived decreasing mass accretion rate with decreasing radius (top right panel of Fig.~\ref{fig:fig7}) could be caused by such an
outflow or by a recent increase of the external mass accretion rate with a characteristic timescale smaller than the viscous timescale
at $R \sim 15$~pc, which is $t_{\rm visc}=3$~Myr or $\sim 20$ dynamical timescales.
We suppose that an outflow also entrains the CRs produced in the SN shocks. In this case, we expect a lower CR injection normalization $k_{\rm CR}$ than
predicted by the Vollmer et al. (2025) model because the model does not include outflows.
Since the normalization of the CR injection rate does not decrease with decreasing radius (Fig.~\ref{fig:fig10b}, we think that
the decreasing mass accretion rate with decreasing radius is caused by a variation of the mass accretion rate at radii $R \ga 30$~pc.
It is thus probable that the mass accretion rate in the inner $\sim 10$~pc of Circinus increases significantly within the next few Myr.

Goesaert et al. (2025) estimated the mass accretion rate within an elongated feature ('the hook') with a high radial velocity of $-43$~km\,s$^{-1}$ located at $R=14$~pc to
be between $0.3$ and $2.2$~M$_{\odot}$yr$^{-1}$. Following Maiolino et al. (2000), the mass accretion rate is given by
\begin{equation}
  \dot{M}=\Sigma\, w\, v_{\rm rad}\ ,
\end{equation}
where $w=1$~pc is the width of the elongated feature.
At $R=5$~pc the gas surface density is $\Sigma \sim 1500$~M$_{\odot}$pc$^{-2}$ (lower middle panel of Fig.~\ref{fig:fig7}) the local mass accretion rate is $\dot{M}=0.07$~M$_{\odot}$yr$^{-1}$. Our estimate of the mass accretion rate is about a factor of two lower (upper right panel of Fig.~\ref{fig:fig7}).

Our mass accretion rate increases monotonically from $\dot{M}=0.04$~M$_{\odot}$yr$^{-1}$ to $\sim 1$~M$_{\odot}$yr$^{-1}$ between $3$ and $14$~pc.
These mass accretion rates are significantly smaller than those of Goesaert et al. (2025) who assumed much higher gas fraction of $0.5$ compared to
our value of $0.03$.

The model gas mass accretion rate within the maser disk ($R \le 1$~pc) is $\dot{M} \sim 0.03$~M$_{\odot}$yr$^{-1}$ (lower right panel of Fig.~\ref{fig:fig7}).
This is still a factor of ten higher than the mass accretion rate onto the supermassive black hole, assuming a bolometric luminosity of
$L_{\rm bol}=3 \times 10^{43}$~erg\,s$^{-1}$ (Moorwood et al. 1996) and leaves ample room for an outflow (Isbell et al. 2022, Goesaert et al. 2025),
which is launched at a radius most probably close to the dust sublimation radius ($R \la 0.1$~pc).

In the following, a relation between the mass outflow rate of the maser disk and that onto the black hole will be derived.
Chan \& Krolik (2017) gave the mass accretion rate onto the black hole in their model of a radiation-driven outflow (their Eq.~45):
\begin{equation}
  \label{eq:eq1}
  \dot{M}_{\rm outflow}=\gamma \big(\frac{L_{\rm UV}}{0.1\,L_{\rm E}}\big)^{\frac{3}{4}} \big(\frac{M_{\rm BH}}{10^7~{\rm M}_{\odot}}\big)^{\frac{3}{4}}~{\rm M}_{\odot}{\rm yr}^{-1}\ ,
\end{equation}
where $L_{\rm UV}$ and $L_{\rm E}$ are the UV and Eddington luminosities and $M_{\rm BH}$ is the mass of the black hole.
The UV luminosity is assumed to be half the bolometric luminosity $L_{\rm bol}$. This yields
\begin{equation}
  \label{eq:eq2}
  L_{\rm UV}=0.5\times 0.1\times \dot{M}_{\rm BH} c^2\ ,
\end{equation}
where $c$ is the speed of light.
The Eddington luminosity is given by
\begin{equation}
  \label{eq:eq3}
  L_{\rm E}=1.26 \times 10^{38} \big(\frac{M_{\rm BH}}{{\rm M}_{\odot}}\big)\ {\rm erg\,s}^{-1}\ .
\end{equation}
Putting together Eqs.~\ref{eq:eq1} to \ref{eq:eq3} we obtained
\begin{equation}
  \dot{M}_{\rm outflow}=10.5\,\gamma\,\dot{M}_{\rm BH}^{\frac{3}{4}}~{\rm M}_{\odot}{\rm yr}^{-1}\ .
\end{equation}
With $\dot{M}$ being the mass accretion rate outside the maser disk, mass conservation between the inflow and outflow yields
\begin{equation}
  \dot{M}=\dot{M}_{\rm BH}+\dot{M}_{\rm outflow}=\dot{M}_{\rm BH}+10.5\,\gamma\,\dot{M}_{\rm BH}^{\frac{3}{4}}\ .
\end{equation}
Since the second term on the right hand side is much larger than the first one, we obtained
\begin{equation}
  \label{eq:macc}
  \dot{M}_{\rm BH} \sim \big(\frac{\dot{M}}{10.5\,\gamma}\big)^{\frac{4}{3}}\ .
\end{equation}
With $\dot{M}=0.04$~M$_{\odot}$yr$^{-1}$ (upper right panel of Fig.~\ref{fig:fig7})
and $\dot{M}_{\rm BH}=5\times 10^{-3}$~M$_{\odot}$yr$^{-1}$ the normalization of Eq.~\ref{eq:eq1} is $\gamma=0.2$.
This normalization is a factor of four lower than that given by Chan \& Krolik (2017; UV+IR outflow).

In the presence of a magnetocentrifugal wind the magnetized outflow extracts angular momentum permitting gas to accrete.
In this case, our model, which is solely based on turbulent viscosity, underestimates the real mass accretion rate. Vollmer et al. (2018)
estimated a mass accretion rate of the thin maser disk with a magnetocentrifugal wind of $\dot{M}=0.12$~M$_{\odot}$yr$^{-1}$ (their Table~2).
In this case $\gamma=0.6$, which is only $40$\,\% smaller than the normalization given by Chan \& Krolik (2017).
We might speculate that the sudden drop of the mass accretion rate and Toomre $Q$ parameter around $R \sim 7$~pc (Fig.~\ref{fig:fig7})
are due to the onset of a magnetocentrifugal wind,
which is responsible for mass accretion. Indeed, the gas disk becomes geometrically thin ($H/R \sim 0.1$) at that radius.

Given the distribution of the free-free emission in the inner $5$~pc (Fig.~6 of Goesaert et al. 2025), it was not surprising that we had to mask
the inner $3$-$4$~pc when we extracted the radial profiles of the continuum emission. However, we would have expected that a masking of
a cone (triangle) in the HR data was enough to remove the free-free emission if the launching point was around $R=0.1$-$0.2$~pc (Goesaert et al. 2025).
We interpret the fact that we had to mask a rectangular region as due to clumpy outflow with an area filling factor of the outflow walls
smaller than unity.
 
We conclude that, contrary to large-scale galactic disks, the energy injection rate into the CND is dominated by the gain of gravitational potential
energy through gas accretion. Since the viscous timescale is shorter than the star formation timescale, the gas distribution and the mass
accretion rate close to the central black hole are set by the variation of the external mass accretion rate and outflows, which can occur at different radii.
We argue in favor of a SN-driven outer outflow from the CND and a radiation-driven inner outflow close to the dust sublimation radius.
In addition, a magnetocentrifugal wind might be present at $R \la 7$~pc.

\subsection{AGN obscuration}

The disk thickness is shown in Fig.~\ref{fig:fig13}.
The disk has a half opening angle of $7.5^{\circ}$ for $R \le 6$~pc.
This is $1.5$ times larger than the half opening angle determined by the modeling of high-resolution mid-infrared
interferomteric observations (Stalevski et al. 2019). Isbell et al. (2022) found a thin disk of $2$~pc diameter with a height $<0.2$~pc in
their high-resolution mid-infrared interferomteric imaging observations, which yields a disk half opening angle $<10^{\circ}$.
We note that these mid-infrared observations have a much higher spatial resolution than the ALMA HR data.
With an inclination angle of $70$-$75^{\circ}$, the gas disk is not responsible for the Compton thickness of the AGN, which implies an obscuring column density higher
than $10^{24}$~cm$^{-2}$. We speculate that the Compton-thick structure is a puff-up of the maser disk created by the AGN radiation close to the
dust sublimation radius (see lower left panel of Fig.~2 of Chan \& Krolik 2017).

\section{Conclusions \label{sec:conclusions}}

The unique ALMA band~6, 7, and 9 dataset of the inner $30$~pc of the Circinus dataset permitted a detailed modeling leading to the extraction of the physical
parameters of the turbulent star-forming gas disk. We applied the model to a set of high-resolution ($2$~pc) and low-resolution ($4$~pc) data
(Table~\ref{tab:data}).
Nine observables ($3$ continuum maps and $6$ CO, HCN, and HCO$^+$ line maps) lead to the
determination of $4$ open model parameters (Toomre $Q$ parameter, gas mass accretion rate $\dot{M}$, metallicity, normalization of the CR injection rate).
The model input parameters are the rotation curve and the radial profiles of the stellar surface density and stellar velocity dispersion.
The model gives access to the gas volume density, star formation rate volume density, gas disk height, turbulent driving length scale, and turbulent velocity dispersion.  
Based on the detailed comparison between observed radial profiles and maps on the one hand and the corresponding models we conclude
\begin{enumerate}
\item
  the model reproduces the observed continuum and molecular line radial profiles (Fig.~\ref{fig:fig2} and \ref{fig:fig5}) and maps
  (Fig.~\ref{fig:fig3} and Fig.~\ref{fig:fig6});
\item
  the model predicts the turbulent velocity of the gas disk (top left panel of Fig.~\ref{fig:fig7});
\item
  the Toomre $Q$ parameter is $Q > 10$. We found $Q=10$ for $3 < R < 7$~pc and $Q=30$ for $R \ga 10$~pc;
  the mass accretion rate decreases from $0.6$~M$_{\odot}$yr$^{-1}$ at $R=15$~pc to $0.04$~M$_{\odot}$yr$^{-1}$ for $R \le 7$~pc (Fig.~\ref{fig:fig7});
\item
  the turbulent and radial velocities and the mass accretion rate are higher, the Toomre $Q$ parameter is lower in the gas arms compared to the interarm regions
  (Fig.~\ref{fig:fig11});
\item
  the CO conversion factor derived from the model is $\alpha_{\rm CO}=0.5$ for $R> 8$~pc and $\alpha_{\rm CO}=1$ for $R \le 8$~pc (Fig.~\ref{fig:fig9});
\item
  the total gas mass of the CND is $2.7 \times 10^6$~M$_{\odot}$, the star formation rate is $\dot{M}_*=0.03$~M$_{\odot}$yr$^{-1}$;
  these values are consistent with those from the literature (Sec.~\ref{sec:ISM});
\item
  gas viscosity determines the evolution of the gas disk ($t_{\rm visc} < t_*$); the viscous timescale at $R=10$~pc is $t_{\rm visc} \sim 10$~Myr;
  the energy injection to maintain turbulence is dominated by the gain of gravitational energy through radial gas accretion for $R<3$~pc and $R>8$~pc
  (Fig.~\ref{fig:fig10}); the galactic disk in the inner $30$~pc of the Circinus galaxy is different from the kpc-scale disk where the viscous timescale is much
  longer than the gas depletion timescale and the energy injection rate to maintain turbulence is dominated by SN explosions;
\item
  based on the low CO conversion factor (Sect.~\ref{sec:ISM}) and the star formation law ($\dot{\Sigma}_* \propto \Sigma^2$; Fig.~\ref{fig:fig12}),
  the inner $30$~pc of the Circinus galaxy can be qualified as a starburst environment;
\item
  the maser disk $R \la 1$~pc can be modeled as a turbulent clumpy star-forming disk (Sect.~\ref{sec:SFR}).
  We found a mass accretion rate of $\dot{M}=0.03$~M$_{\odot}$yr$^{-1}$ in the maser disk. In the presence of a magnetocentrifugal wind it can be a factor of three higher;
  The relation between the mass accretion within the maser disk and that onto the black hole is given by Eq.~\ref{eq:macc} with $\gamma=0.2$-$0.7$;
\end{enumerate}

Since the evolution of the gas disk is governed by the viscous timescale, the gas surface density distribution in the inner $30$~pc of the
Circinus galaxy is set by the evolution of the external mass accretion rate at $R > 30$~pc. In a steady state configuration with a constant external accretion rate
over a time longer than the disk's viscous timescale, we expect a radially constant mass accretion rate if no outflow is present. Since this is not observed, the external mass
accretion rate presumably varies within a timescale shorter than $\sim 10$~Myr. 
Indeed, if the external mass accretion rate is increased by an asymmetric oval gravitational disk potential inducing strong gas arms,
we expect that a variation of the external mass accretion rate
occurs on a timescale of several dynamical times at $R \sim 100$~pc, which corresponds to several Myr. The infalling gas symmetrizes at $R \sim 30$~pc and forms the CND. 
The CND is at present in a phase of high external mass accretion rate caused by the three prominent gas arms highlighted by Izumi et al. (2018).
The ISM of the Circinus galaxy is thus driven inwards by nested spiral structures from kpc (Lasrado et al. 2025) to pc scales (Goesaert et al. 2025).

\clearpage

\begin{appendix}

\section{The model gas disk \label{sec:dgasdisk}}

In this section and the following, we briefly describe the Vollmer et al. (2017)  model with the minor modifications given in Vollmer et al. (2025).
The meaning of the different variables is given in Tables~\ref{tab:mparams} and \ref{tab:parameters}.

\begin{table*}
\begin{center}
\caption{Model Parameters (see also Table~\ref{tab:mparams}).\label{tab:parameters}}
\begin{tabular}{lll}
\hline\hline
Parameter & Unit & Explanation \\
\hline
$G=5 \times 10^{-15}$ & pc$^{3}$yr$^{-1}$M$_{\odot} ^{-1}$ & gravitation constant \\
$R$ & pc & galactocentric radius \\
$\kappa$ & yr$^{-1}$ & epicyclic frequency \\
$\Omega=v_{\rm rot}/R$ & yr$^{-1}$ & angular velocity \\
$\Phi_{\rm V}$ & & volume filling factor \\\
$l_{\rm cl}$ & pc & cloud size \\
$\rho_{\rm cl}=\rho/\Phi_{\rm V}$ & M$_{\odot}$pc$^{-3}$ & cloud density \\
$\xi=9.2 \times 10^{-8}$ & pc$^2$yr$^{-2}$ & constant relating SN energy input to star formation  \\
$v_{\rm r}$ & pc\,yr$^{-1}$ & gas radial velocity \\
$\nu = v_{\rm turb} l_{\rm driv}$ & pc$^{2}$yr$^{-1}$ & gas viscosity \\
$\alpha$ & yr\,M$_{\odot}$pc$^{-3}$ & constant of molecule formation timescale \\
${\rm SFE}=\dot{\Sigma}_{*}/\Sigma$ & yr$^{-1}$ & star formation efficiency \\
$t_{\rm ff}^{l}$ & yr & cloud free fall timescale at size $l$ \\
$t_{\rm turb}^{l}$ & yr & cloud turbulent timescale at size $l$ (turbulent crossing time) \\
\hline
\end{tabular}
\end{center}
\end{table*}

In the model, the disk scale height is determined unambiguously by the assumption of hydrostatic equilibrium and the turbulent pressure
(Elmegreen 1989):

\begin{equation}
p_{\rm turb}=\rho v_{\rm  turb}^{2} = \frac{\pi}{2} G \Sigma ( \Sigma + \Sigma_{*} \frac{v_{\rm turb}}{v_{\rm disp}^{*}})~,
\label{eq:pressure}
\end{equation}

\noindent where $\rho$ is the large-scale density, $v_{\rm turb}$ the gas turbulent velocity in the disk, $v_{\rm disp}^{*}$ the stellar vertical velocity dispersion, 
and $\Sigma$ the surface density of gas and stars.

Following Vollmer \& Leroy (2011) and Vollmer et al. (2021), the SFR per unit volume is given by
\begin{equation}\label{eq:star-form}
\dot{\rho}_{*} = \Phi_{\rm V} \frac{\rho}{t_{\rm ff}^{\rm l}}\ .
\end{equation}
where $\Phi_{\rm V}$ is the volume filling factor, $l$ and $\rho_{\rm l}=\rho/\Phi_{\rm V}$ the size and density of selfgravitating clouds,
and $t_{\rm ff}^{\rm l}=\sqrt{3\pi/(32\,G\,\rho_{\rm l})}$ the local free-fall time.
This prescription leads to an SFR per unit area which is proportional to the gas pressure $p=\rho v_{\rm turb}^2$ (Fig.~A.1 of Vollmer et al. 2022)
as expected if star formation is pressure-regulated and feedback-modulated (Ostriker \& Kim 2022).

For self-gravitating clouds with a virial parameter of two (Bertoldi \& McKee 1992), the turbulent crossing time $t_{\rm turb,cl}$ equals
$1.4$ times the free-fall time
$t_{\rm turb,cl}$:
\begin{equation}
t_{\rm turb,cl} =  \frac{l_{\rm cl}}{2\,v_{\rm turb,cl}} = 1.4 \times t_{\rm ff,cl}=1.4 \times \sqrt{\frac{3 \pi \Phi_{\rm V}}{32 G \rho}}
\label{eq:tturbcl}
\end{equation}
where $l_{\rm cl}$ and $v_{\rm turb,cl}$ are the size and the turbulent 1D velocity dispersion of a single gas cloud, respectively. Following Larson's
law (Larson 1981), we can simplify the expression of the turbulent crossing time:
\begin{equation}
t_{\rm turb,cl}=\frac{l_{\rm cl}}{2\,v_{\rm turb,cl}} = \frac{l_{\rm driv}}{2\,v_{\rm turb} \sqrt{\delta}}\ .
\end{equation}
where $\delta$ is the scaling between the driving length scale and the
size of the largest self-gravitating structures, such as $\delta = l_{\rm driv}/l_{\rm cl}$.
The expression for the SFR then becomes
\begin{equation}
  \label{eq:sfrrec}
\dot{\rho}_*= 2.8\,\sqrt{\delta} \Phi_{\rm V} \rho \frac{v_{\rm turb}}{l_{\rm driv}}\ .
\end{equation}
and $\dot{\Sigma}_*=\dot{\rho}_* l_{\rm driv}$.

The SFR density is used to calculate the rate of energy injection by supernovae (SNe):
\begin{equation}
  \label{eq:energyflux}
  \frac{\dot{E}_{\rm SN}}{\Delta A}=\xi\,\dot{\Sigma}_{*} 
  = \xi\,\dot{\rho}_{*} l_{\rm driv}=\frac{1}{2} \Sigma \nu \frac{v_{\rm turb,3D}^{2}}{l_{\rm driv}^{2}}
  = \frac{3\,\sqrt{3}}{2} \Sigma \frac{v_{\rm turb}^3}{l_{\rm driv}}\ ,
\end{equation}
where $\Delta A$ is the unit surface element of the disk, $\dot{\Sigma}_{*}$ and $\dot{\rho}_{*}$ the star formation per unit area and unit volume,
$\Sigma$ the total gas surface density, $\nu$ the turbulent viscosity, $v_{\rm turb}$ the turbulent velocity of the gas, and the CO disk thickness
is assumed to be $l_{\rm driv}$. The turbulent viscosity is $\nu=\sqrt{3}v_{\rm turb} l_{\rm driv}$.
The energy injection rate is related to the turbulent
velocity dispersion and the driving scale of turbulence. These quantities in turn provide estimates of the clumpiness of gas in the disk (i.e., the
contrast between local and average density) and the rate at which viscosity moves matter inward.
The derived expressions for $\Phi_{\rm V}$ and $\dot{\Sigma}_*$ are
\begin{equation}
\Phi_{\rm V}=\frac{4.1\,G\,\rho\,l_{\rm driv}^2}{3\,\pi\,\delta\,v_{\rm turb}^2}
\end{equation}
and
\begin{equation}
\dot{\Sigma}_*=\frac{11.4\,G\,\rho^2 l_{\rm driv}^2}{3\,\pi\,\sqrt{\delta} v_{\rm turb}}\ .
\label{eq:sigmastar1}
\end{equation}

Following Vollmer et al. (2017), our leaky-box model uses an effective yield, meaning that enriched gas can escape and be accreted from the circumgalactic medium.
This leads to a gas metallicity of
\begin{equation} \label{eq:zzodot}
Z/Z_{\odot}=k_{\rm Z}\,(0.61~{\rm yr\,M_{\odot}pc^{-3}})/\alpha\ ,
\end{equation}
where $k_{\rm Z}$ is the normalization factor and
\begin{equation}
\label{eq:metulirgs}
\alpha=\big( \ln(\frac{\Sigma_{*}+\Sigma}{\Sigma})\big)^{-1}\ {\rm yr\,M_{\odot}pc^{-3}}\ .
\end{equation}

The Toomre parameter of the gas is given by
\begin{equation}
  Q=\frac{v_{\rm turb}\kappa}{\pi\,G\,\Sigma}\ ,
\end{equation}
where $\kappa$ is the epicyclic frequency.
Since a gas disk with $Q<1$ is gravitationally unstable, a model with $Q_{\rm crit} \sim 1$ can be seen as a disk with maximum gas content.
Within our model framework $Q$ can be interpreted as a measure of the gas mass (see also Eq.~\ref{eq:Q1}) for a given rotation curve.
We acknowledge that disk thickness can modify $Q_{\rm crit}$. In thick discs, gravitational forces are weaker, and as a consequence,
the $Q_{\rm crit}$ threshold can drop below unity.

The turbulent velocity can be driven by SN explosions or external gas infall, which stir the disk and lead to viscous transport of
angular momentum. In addition, star formation or outflows might remove gas from the viscous evolution.
Following Lin \& Pringle (1987), the evolution of the gas surface density is given by
\begin{equation}
\frac{\partial \Sigma}{\partial t}=-\frac{1}{R}\frac{\partial}{\partial R}\left(
\frac{(\partial/\partial R)[\nu \Sigma R^3 ({\rm d}\Omega/{\rm }dR)]}{({\rm d}/{\rm d}R)(R^2 \Omega)}\right)
-\dot{\Sigma}_{*}-\dot{\Sigma}_{\rm out}+\dot{\Sigma}_{\rm sph}\ ,
\label{eq:linpringle}
\end{equation}
where $\nu$ is the gas viscosity and $\Omega$ the angular velocity.  
Gas accreting from the halo gives rise to the spherical mass accretion rate $\dot{\Sigma}_{\rm sph}$.
Supernova-driven or magneto-rotational disk outflows decrease the gas surface density by $\dot{\Sigma}_{\rm out}$.
 
By approximating $\partial/\partial R \sim 1/R$, the global viscous evolution becomes
\begin{equation}
\frac{\partial \Sigma}{\partial t} \sim \frac{\Sigma \nu}{R^2}-\dot{\Sigma}_{*}-\dot{\Sigma}_{\rm out}+\dot{\Sigma}_{\rm sph}=\Sigma\,(t_{\rm visc}^{-1}+t_*^{-1}+t_{\rm out}^{-1})+\dot{\Sigma}_{\rm sph}\ ,
\label{eq:parsigma}
\end{equation}
where the viscous timescale is $t_{\rm visc}=R/v_{\rm r}=R^2/\nu$ with  the radial velocity $v_{\rm r}$, and the gas viscosity $\nu=v_{\rm turb}\,l_{\rm driv}$.
The local disk mass accretion rate at a given radius $R$ is given by
\begin{equation}
  \dot{M}(R)=2\,\pi\,R\,\Sigma(R)\,v_{\rm r}(R)=2\,\pi\,\nu(R)\,\Sigma(R)\ .
\end{equation}
The local viscous timescale then is
\begin{equation}
  t_{\rm visc}(R)=\frac{2\,\pi\,R^2 \Sigma(R)}{\dot{M}(R)}\ .
  \label{eq:tvisc1}
\end{equation}

For $t_{\rm visc} \ll t_*$, $t_{\rm visc} \ll t_{\rm out}$, and $\dot{\Sigma}_{\rm sph} \ll \Sigma/t_{\rm visc}$ within a steady state disk
($ \frac{\partial \Sigma}{\partial t}=0$), one obtains
\begin{equation}
\frac{1}{R}\frac{\partial}{\partial R}\left(\frac{(\partial/\partial R)[\nu \Sigma R^3 ({\rm d}\Omega/{\rm }dR)]}{({\rm d}/{\rm d}R)(R^2 \Omega)}\right)=0\ ,
\end{equation}
which leads to the classical steady-state accretion disk with a constant mass accretion rate (Pringle 1981).
Contrary to the classical model, the disk mass accretion, which is only caused by turbulent viscosity, varies with galactic radius within our model prescription.
Gas inflow and outflow together with the viscous radial gas transport set the gas surface density (Eq.~\ref{eq:linpringle}).

The model relies on several empirical calibrations: the relation between the stellar
velocity dispersion and the stellar disk scale length, the relationship between the star-formation rate (SFR) and the energy injected into
the ISM by SNe, and the characteristic time of H$_2$ formation, which is related to the gas metallicity and the gas density.
The fraction between the molecular and the total gas mass is governed by the turbulent crossing time $t^{\rm l}_{\rm turb}$ and the molecule formation timescale.
In addition, photodissociation of molecules is taken into account.

The model inputs are the observed rotation curve and stellar surface density. The stellar velocity dispersion has to be assumed.
These parameters define the gravitational potential of the disk.
The deeper the potential the higher is the gas density if the gas velocity dispersion is constant (Eq.~\ref{eq:pressure}).
Equations~A25--A32 of Vollmer \& Leroy (2011) give the dependencies of the disk properties on the model free parameters
$Q$ and $\dot{M}$, the observed stellar surface density $\Sigma_*$, and the observed angular velocity $\Omega=v_{\rm rot}/R$:
the gas velocity dispersion slowly increases with $Q$ and $\dot{M}$
\begin{equation}
  v_{\rm turb} \propto Q^{0.23} \dot{M}^{0.15} \Sigma_*^{0.15} \Omega^{-0.15}\ .
\end{equation}
The gas surface density mostly depends on the Toomre $Q$ parameter and the angular velocity
\begin{equation}
  \Sigma \propto Q^{-0.85} \dot{M}^{0.23} \Sigma_*^{0.15} \Omega^{0.85}\ .
  \label{eq:Q1}
\end{equation}
The star formation surface density varies approximately as
\begin{equation}
  \dot{\Sigma}_* \propto Q^{-1.08} \dot{M}^{0.38} \Sigma_*^{0.92} \Omega^{1.08}\ ,
  \label{eq:sfr11}
\end{equation}
the turbulent driving length scale as
\begin{equation}
l_{\rm driv} \propto Q^{0.69} \dot{M}^{0.54} \Sigma_*^{-0.31} \Omega^{-0.69}\ ,
\end{equation}
the star formation efficiency as
\begin{equation}
  {\rm SFE}\,\propto Q^{-0.23} \dot{M}^{0.15} \Sigma_*^{0.78} \Omega^{0.23}\ ,
\end{equation}
  and the gas viscosity as
\begin{equation}
  \nu \propto Q^{0.85} \dot{M}^{0.77} \Sigma_*^{-0.15} \Omega^{-0.85}\ .
\end{equation}
Increasing $Q$ leads to a higher velocity dispersion, a lower gas and star formation surface density, a larger turbulent driving length scale,
and a lower star formation efficiency. On the other hand, a higher $\dot{M}$ leads to a higher velocity dispersion, a higher star formation
surface density, a larger driving length scale, and a higher star formation efficiency.
A high angular velocity $\Omega$ leads to high gas and star formation surface densities and low driving length scale and gas viscosity.
A high stellar surface density leads to a high star formation surface density and SFE. Since $\Omega$ and $\Sigma_*$ are very high in the inner
$30$~pc of the Circinus galaxy, the gas mass and star formation rate within this region can be quite high even if $Q$ is large. 

The model is divided into a large-scale and a small-scale part. The former yields the surface density, turbulent velocity, disk
height, and gas viscosity. The latter is relevant for self-gravitating gas clouds $(t^{\rm l}_{\rm ff} \le t^{\rm l}_{\rm turb})$.
The non-self-gravitating and self-gravitating clouds obey different scaling relations, which are set by observations.
For each gas density, the mass fraction of clouds of this density is characterized by a log-normal probability distribution function and the Mach number
(Padoan et al. 1997). The temperatures of the gas clouds are calculated via the equilibrium between turbulent mechanical and
CR heating and gas cooling via CO and H$_2$ line emission. The abundances of the different molecules are determined using the time-dependent
gas-grain code Nautilus (Hersant et al. 2009).
This code computes the abundances of chemical species (atoms and molecules) as a function of time by solving the rate equations for a
network of reactions. For gas-phase reactions, we use the kida.uva.2014 network (Wakelam et al. 2015\footnote{the network is available online on the
KIDA website {\rm https://kida.astrochem-tools.org/}}) comprising $489$ species and $6992$ reactions. For grain surface reactions, we use the
desorption, diffusion, activation barrier energies along with a set of grain surface reactions, all from the KIDA database. Both, thermal and
non-thermal desorption processes are taken into account, the latter consisting mainly of CR induced desorption following the formalism
presented Hasegawa et al. (1993).
The model parameters are the time, density, gas temperature, grain temperature, cosmic ray ionization rate, and the elemental
abundances of the elements C, O, and N (C/H$=1.7 \times 10^{-4}$, O/H$=2.4 \times 10^{-4}$, N/H$=6.2 \times 10^{-5}$). 
Grids of models were obtained by varying the cloud life time (20 log spaced steps between $10^3$ and $10^8$~yr), the cloud density (20 log spaced steps
between $10^3$ and $10^8$~cm$^{-3}$) and cloud gas temperatures (20 log spaced steps $10$
and $800$~K). For each type of clouds, the CO, HCN, and HCO+ abundances were interpolated on the grid given the life time, density, and
temperature of the cloud. 

In addition, we take into account the photodissocation of H$_2$ and CO by the interstellar radiation field (Sect.~2.4.2 of Vollmer et al. 2017).
The dust temperatures are calculated via the equilibrium between heating by the interstellar UV and optical
radiation field and cooling via IR emission.

Within our model approach, the free parameters of the model are set by the radial profiles of the molecular line and continuum emission.
These free parameters are the Toomre $Q$ parameter, the local disk mass accretion rate $\dot{M}$ and the normalizations of the gas metallicity $k_{\rm Z}$
and the CR ionization rate ($k_{\rm CR}$). The local disk properties ($\Sigma$, $\dot{\Sigma}_*$, $\rho$, $H$, $v_{\rm turb}$,
$l_{\rm driv}$, $\nu$), are calculated for each galactic radius independently using these parameters.
The model provides the present state of the accretion disk even if this state is set by strong inflows and outflows (Eq.~\ref{eq:parsigma}).
However, the model only yields the local turbulent viscous mass accretion rate (Eq.~\ref{eq:tvisc1}). It is agnostic about the inflow and outflow rates.

\clearpage

\section{Rotation curve}

\begin{figure}
  \centering
  \resizebox{\hsize}{!}{\includegraphics{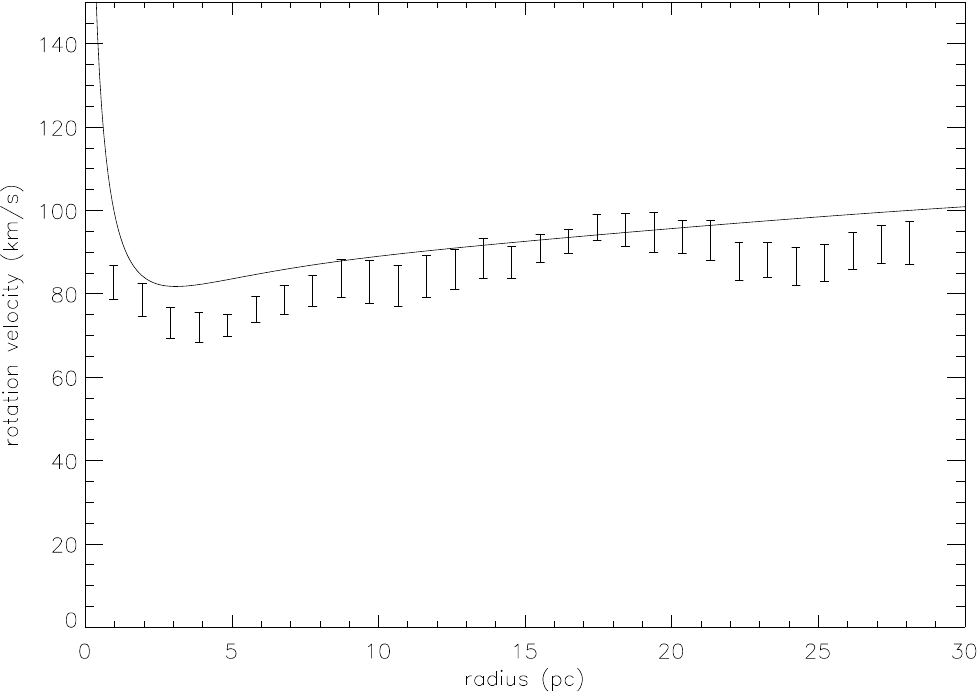}}
  \caption{Model rotation curve (solid line) together with the observed HCN(3-2) rotation curve (Goesaert et al. 2025).
  \label{fig:rotcurve}}
\end{figure}

\section{Metallicity and CR factor}

\begin{figure}
  \centering
  \resizebox{\hsize}{!}{\includegraphics{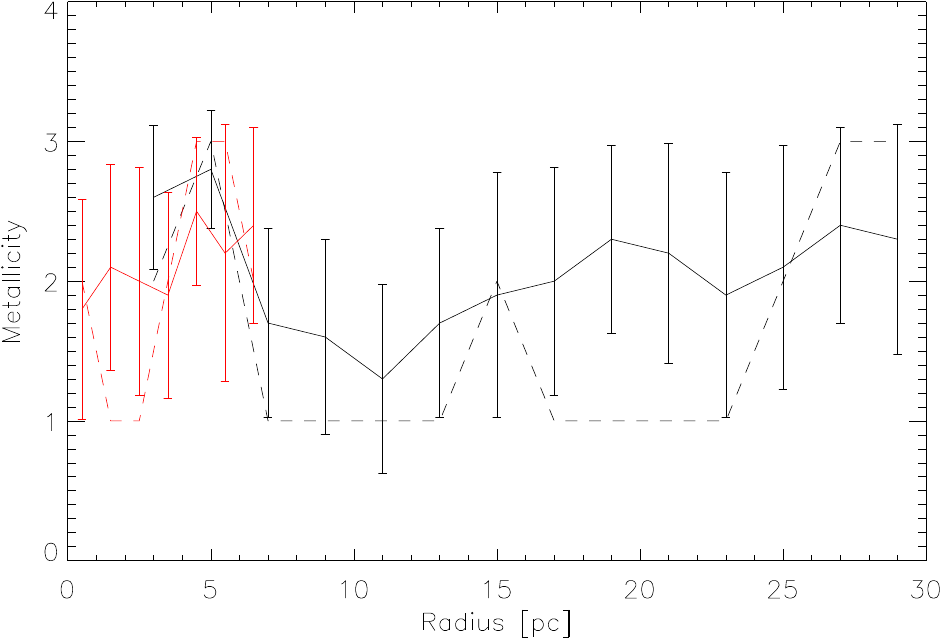}}
  \resizebox{\hsize}{!}{\includegraphics{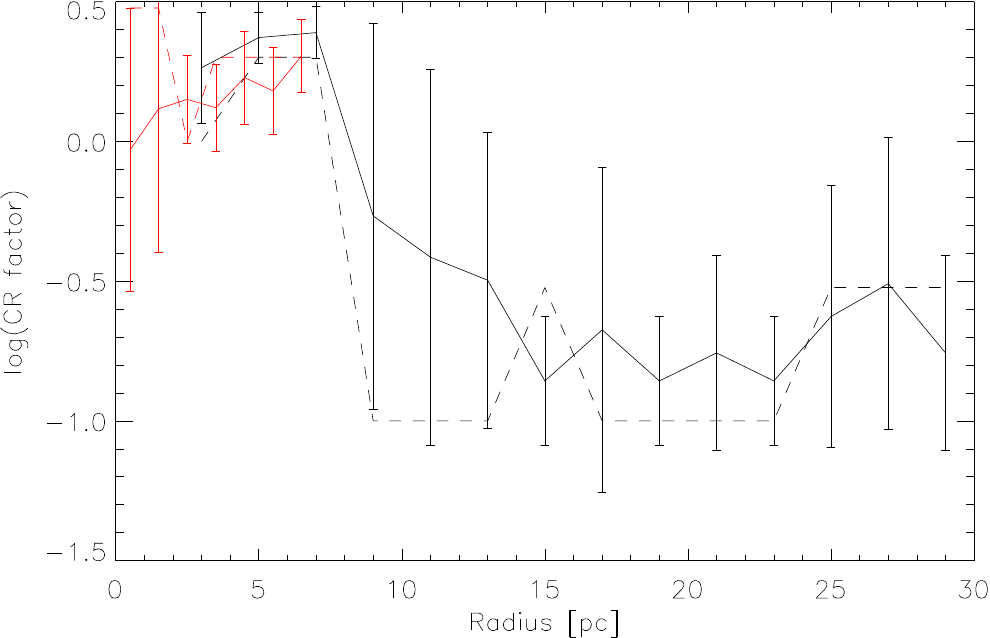}}
  \caption{Radial profiles of the model metallicity normalization $k_{\rm Z}$ (upper panel) and CR injection normalization $k_{\rm CR}$ (lower panel).
    Red lines: HR model; black lines LR model.
  \label{fig:fig10a}}
\end{figure}

\begin{figure}
  \centering
  \resizebox{\hsize}{!}{\includegraphics{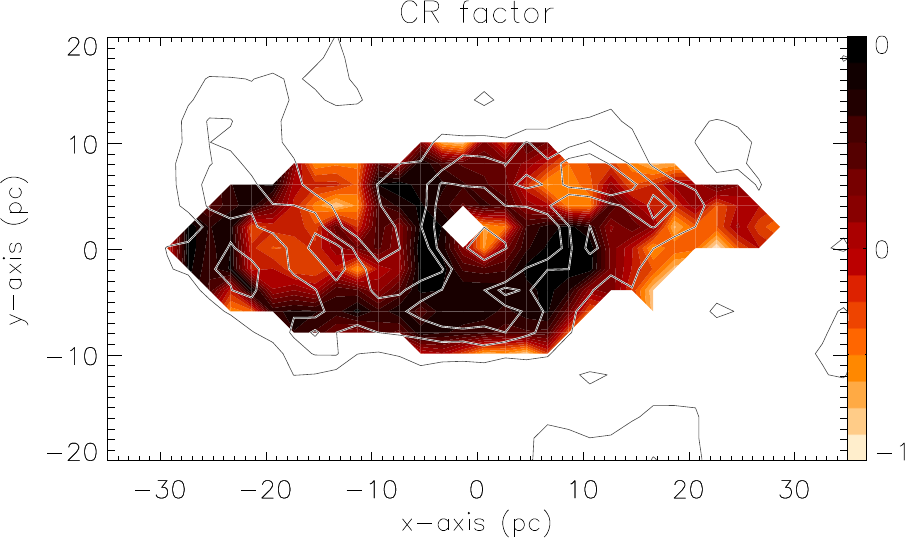}}
  \caption{Map of the model CR injection normalization $k_{\rm CR}$. The HCO$^+$ (4-3) emission is shown by contours to guide the eye.
    \label{fig:fig10b}}
\end{figure}

\section{Nuclear Star cluster}

\begin{figure}
  \centering 
  \resizebox{\hsize}{!}{\includegraphics{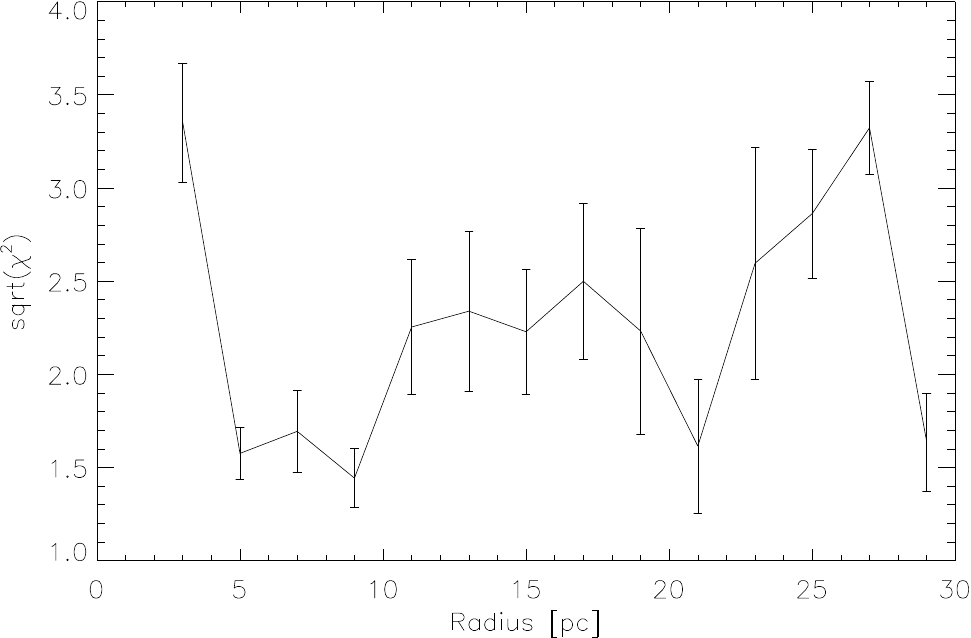}}
  \resizebox{\hsize}{!}{\includegraphics{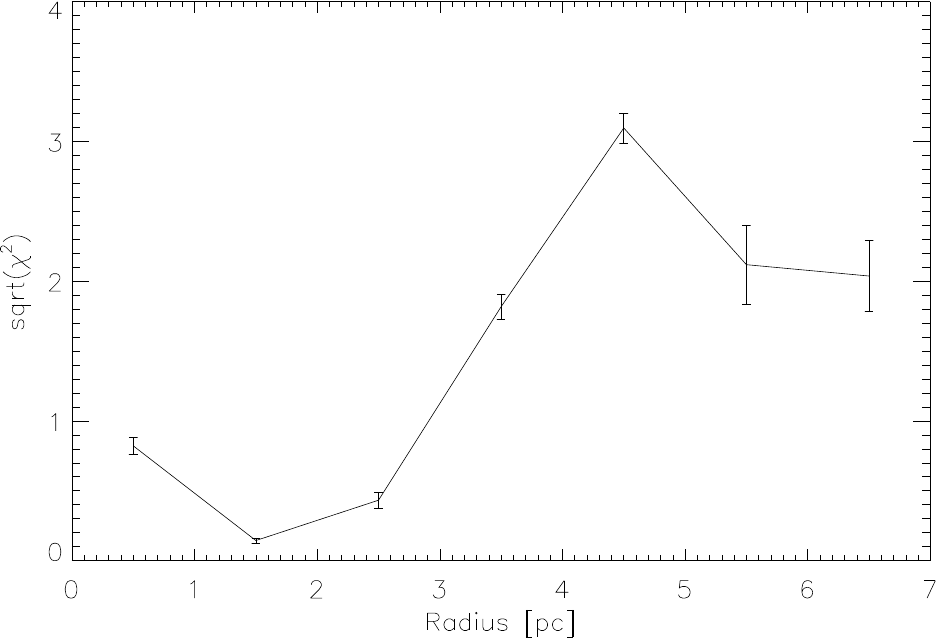}}
  \caption{Mean $\chi^2$ for the $10$ best models with a nuclear stellar cluster and its scatter. Upper panel: LR model, lower panel: HR model.
  \label{fig:khideux_log_nd_pb6_pco32HR_konrad_nooutflow_2disks_18a}}
\end{figure}

\begin{figure*}
  \centering
  \resizebox{\hsize}{!}{\includegraphics{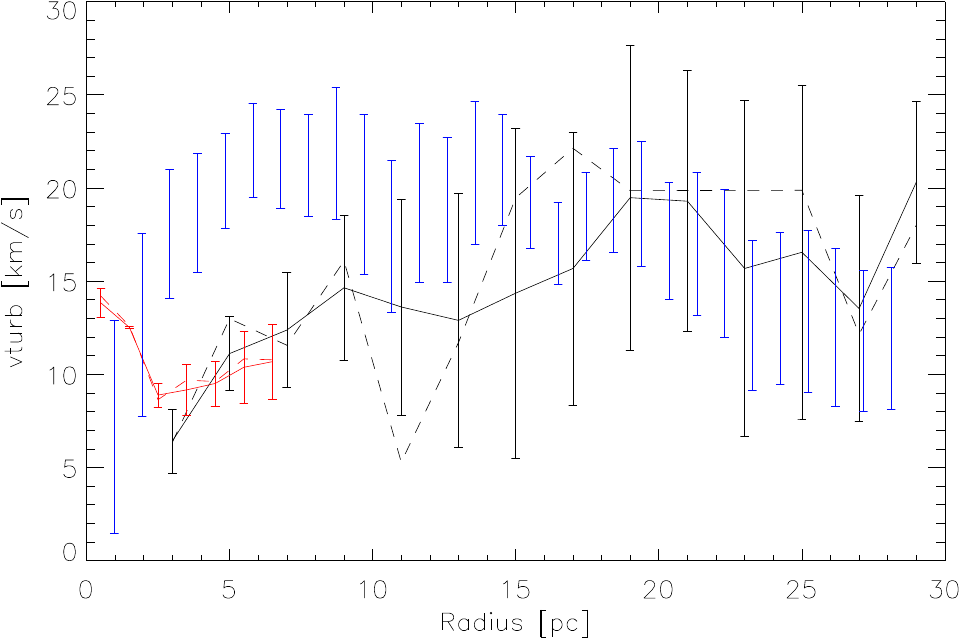}\includegraphics{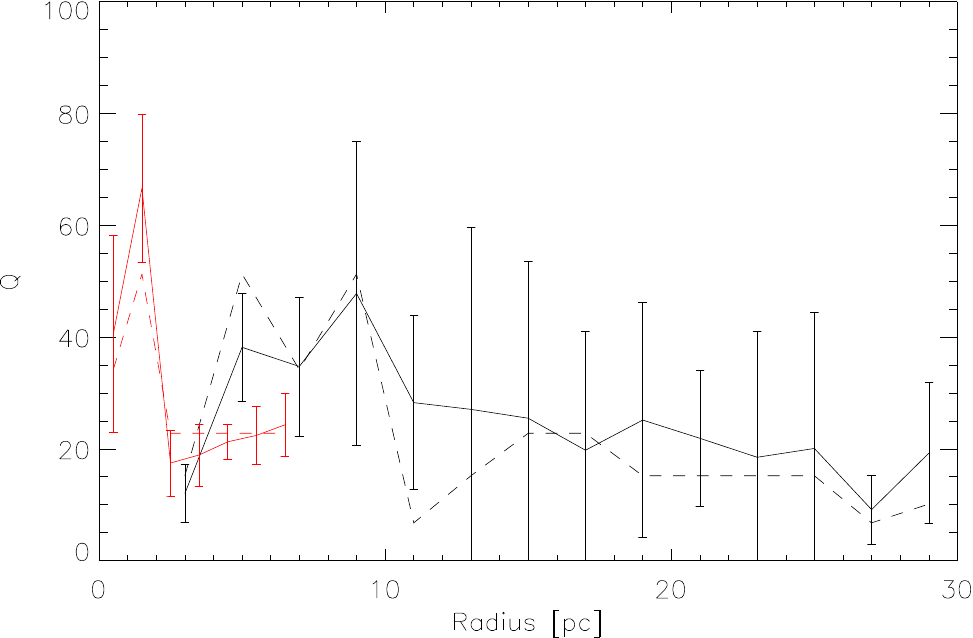}\includegraphics{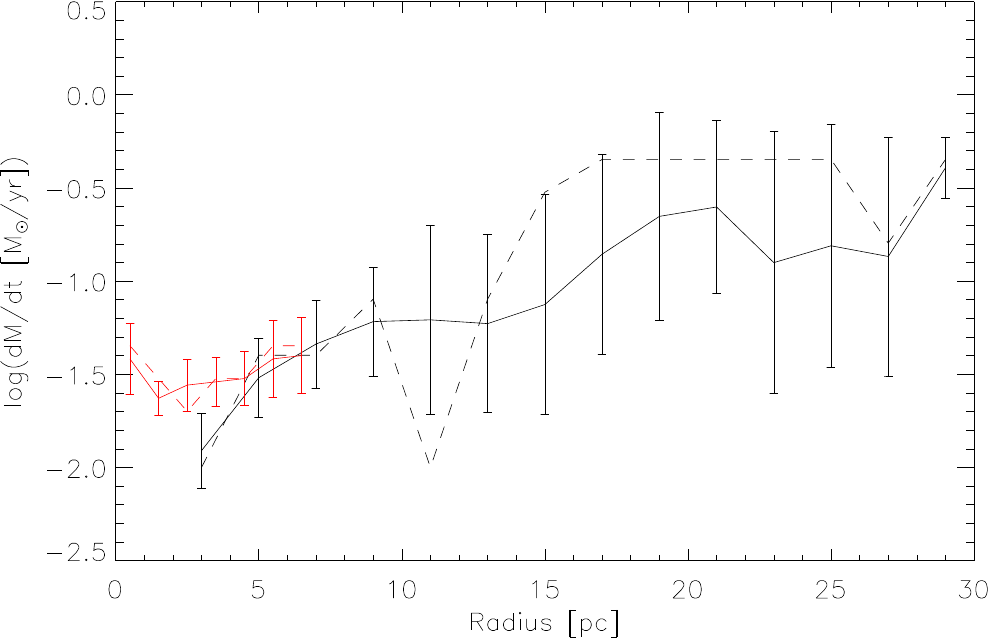}}
  \resizebox{\hsize}{!}{\includegraphics{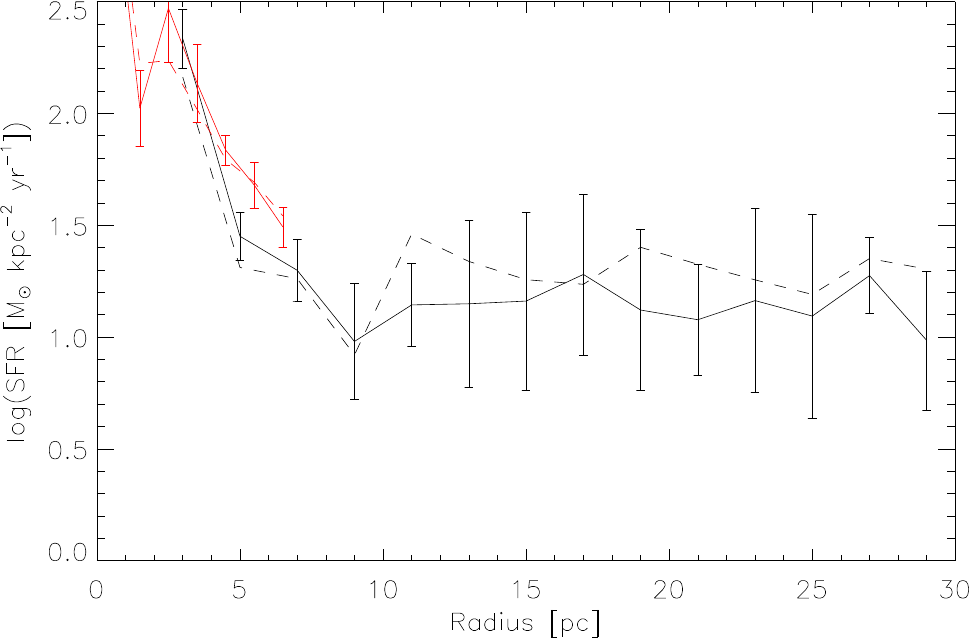}\includegraphics{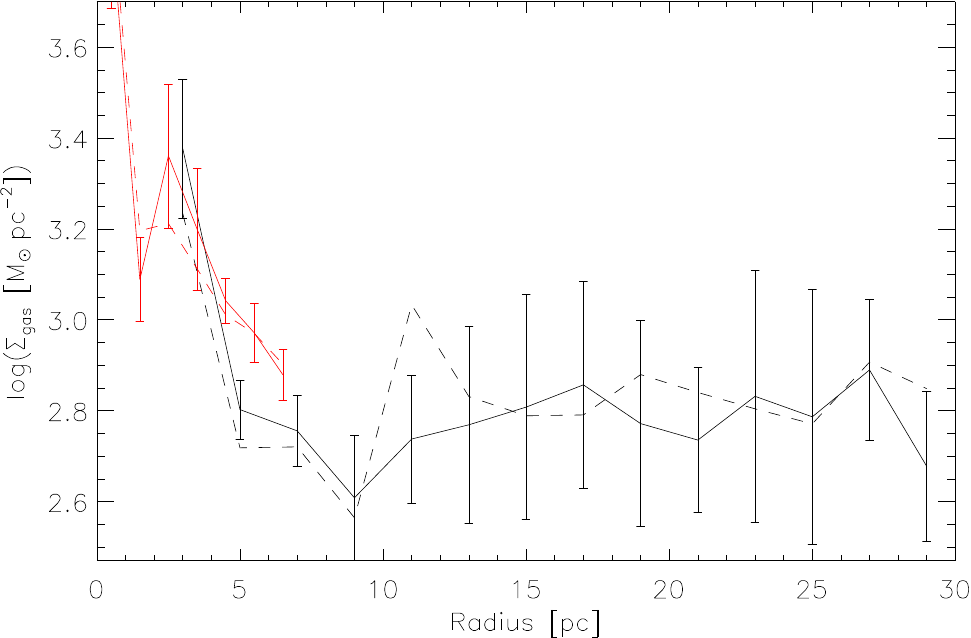}\includegraphics{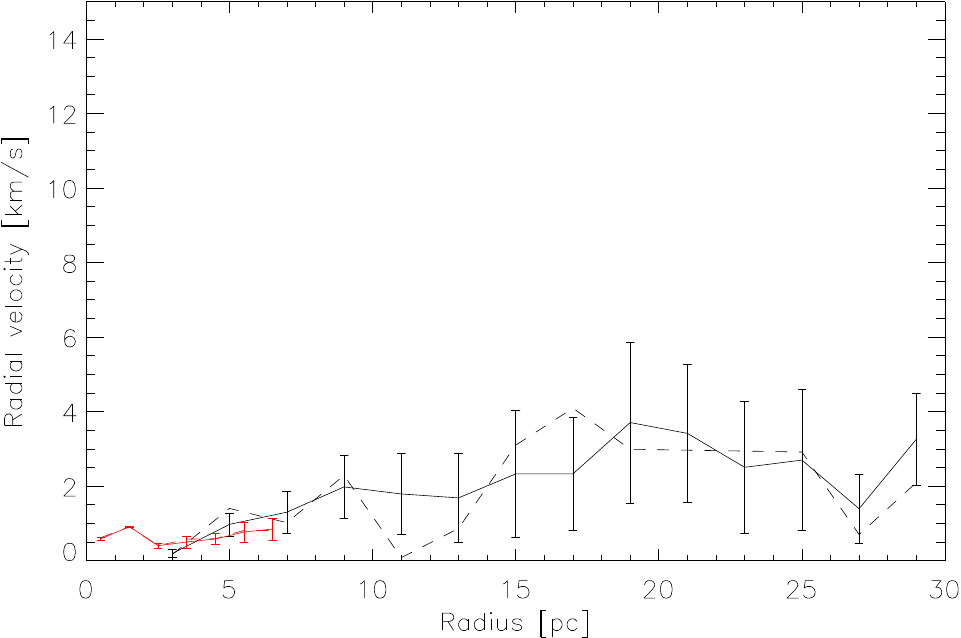}}
  \caption{Radial profiles of the derived physical quantities for the model including a nuclear star cluster. From upper left to lower right: gas velocity dispersion (blue error bars: observed HCN(3-2) velocity dispersion from Goesaert et al. 2025), Toomre $Q$ parameter, mass accretion rate $\dot{M}$, star formation rate surface density, gas surface density, gas radial velocity. Red lines: HR model; black lines LR model.
    \label{fig:fig7a}}
\end{figure*}

\section{Star formation law}

\begin{figure}
  \centering
  \resizebox{\hsize}{!}{\includegraphics{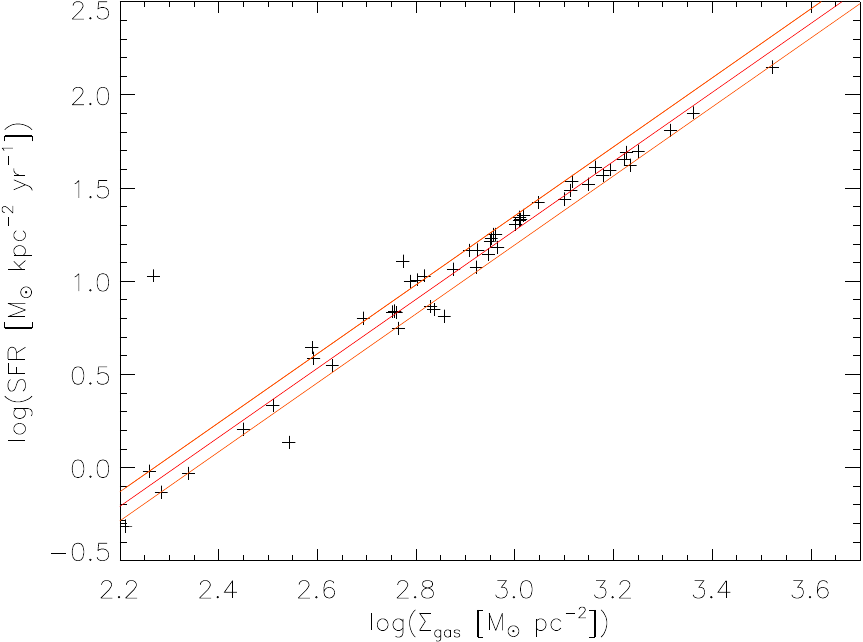}}
  \caption{Star formation surface density as a function of the gas surface density of the LR model. The red solid lines correspond to an outlier-resistant
    linear fit and its scatter. The slope of the log-log relation is $1.9$.
  \label{fig:fig12}}
\end{figure}

\section{Disk height}

\begin{figure}
  \centering
  \resizebox{\hsize}{!}{\includegraphics{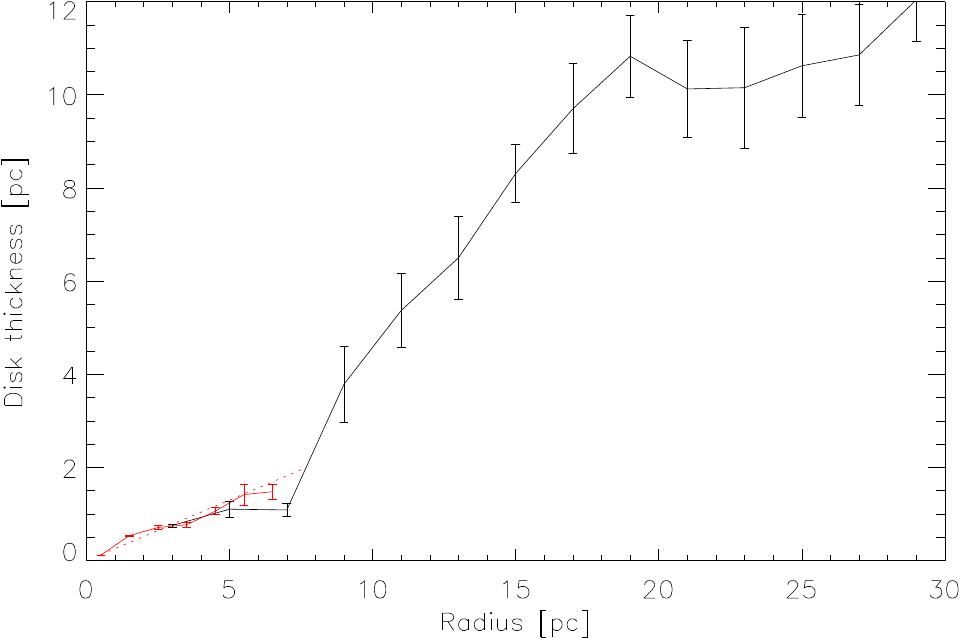}}
  \caption{Disk thickness as a function of the radius for the HR (red) and LR (black) models. Dotted line: linear relation with a slope of $0.26$.
  \label{fig:fig13}}
\end{figure}

\end{appendix}

\end{document}